\documentclass[prb,aps,twocolumn,superscriptaddress,longbibliography]{revtex4-2}

\usepackage[T1]{fontenc}
\usepackage[latin9]{inputenc}
\usepackage{amssymb}
\usepackage{graphicx}
\usepackage{amsmath,color}
\usepackage{mathrsfs}
\usepackage{float}
\usepackage{indentfirst}
\usepackage{subfigure}
\usepackage{multirow}
\usepackage{tabu}
\usepackage{threeparttable}
\usepackage{booktabs}
\usepackage{txfonts}
\usepackage{amsmath,amssymb,bbm}
\usepackage[normalem]{ulem}
\usepackage[euler]{textgreek}
\usepackage{bm}% bold math

\usepackage[dvipsnames]{xcolor}
\usepackage{soul}
\usepackage[colorlinks=true,urlcolor=MidnightBlue,citecolor=MidnightBlue,linkcolor=MidnightBlue]{hyperref}

\newcommand{\mathe}{\mathrm{e}}

\newcommand{\<}{\langle}
\renewcommand{\>}{\rangle}

\newcommand{\tr}{\mathrm{Tr}}

\newcommand\redsout{\bgroup\markoverwith{\textcolor{red}{\rule[0.5ex]{2pt}{0.4pt}}}\ULon}

\newcommand{\LL}{\mathrm{LL}}
\newcommand{\RR}{\mathrm{RR}}
\newcommand{\LR}{\mathrm{LR}}
\newcommand{\rmR}{\mathrm{R}}
\newcommand{\rmL}{\mathrm{L}}

\newcommand{\psiL}{\psi^{\mathrm{L}}}
\newcommand{\psiR}{\psi^{\mathrm{R}}}
\newcommand{\calA}{\mathcal{A}}
\newcommand{\calB}{\mathcal{B}}
\newcommand{\rhoLL}{\rho_{\calA}^{\mathrm{LL}}}
\newcommand{\rhoRR}{\rho_{\calA}^{\mathrm{RR}}}
\newcommand{\rhoLR}{\rho_{\calA}^{\mathrm{LR}}}
\newcommand{\calG}{\mathcal{G}}
\newcommand{\calT}{\mathcal{T}}
\newcommand{\calP}{\mathcal{P}}
\newcommand{\calU}{\mathcal{U}}
\newcommand{\Uone}{\mathrm{U}(1)}

\begin{document}

\preprint{APS/123-QED}

\title{Matrix product state fixed points of non-Hermitian transfer matrices}

\author{Wei Tang}
\affiliation{Department of Physics and Astronomy, Ghent University, Krijgslaan 281, 9000 Gent, Belgium}
\author{Frank Verstraete}
\affiliation{Department of Applied Mathematics and Theoretical Physics, University of Cambridge, Wilberforce Road, Cambridge, CB3 0WA, United Kingdom}
\affiliation{Department of Physics and Astronomy, Ghent University, Krijgslaan 281, 9000 Gent, Belgium}
\author{Jutho Haegeman}
\affiliation{Department of Physics and Astronomy, Ghent University, Krijgslaan 281, 9000 Gent, Belgium}

\date{\today}

\begin{abstract}
    The contraction of tensor networks is a central task in the application of tensor network methods to the study of quantum and classical many body systems.
    In this paper, we investigate the impact of gauge degrees of freedom in the virtual indices of the tensor network on the contraction process, specifically focusing on boundary matrix product state methods for contracting two-dimensional tensor networks.
    We show that the gauge transformation can affect the entanglement structures of the eigenstates of the transfer matrix and change how the physical information is encoded in the eigenstates, which can influence the accuracy of the numerical simulation. 
    We demonstrate this effect by looking at two different examples. 
    First, we focus on the local gauge transformation, and analyze its effect by viewing it as an imaginary-time evolution governed by a diagonal Hamiltonian. 
    As a specific example, we perform a numerical analysis in the classical Ising model on the square lattice.
    Second, we go beyond the scope of local gauge transformations and study the antiferromagnetic Ising model on the triangular lattice.
    The partition function of this model has two tensor network representations connected by a nonlocal gauge transformation, resulting in distinct numerical performances in the boundary matrix product state calculation.  
\end{abstract}

\maketitle

%\tableofcontents

\section{Introduction} 

Tensor networks~\cite{schollwock-density-2011,orus-practical-2014,cirac-matrix-2021,xiang-density-book-2023} have been widely used in the study of quantum and classical many body physics. 
In the context of quantum many body physics, tensor network states, such as matrix product states (MPS), projected entangled-pair states (PEPS), and the multiscale entanglement renormalization ansatz (MERA), have been proven to be a powerful ansatz for representing low-energy states of local Hamiltonians, owing to their unique entanglement characteristics.
Moreover, tensor networks can also be used to represent the partition functions of classical statistical models and the Euclidean path integrals of quantum systems.

When employing the tensor network as a computational tool, one paramount task is the contraction of the tensor network. 
Generally speaking, exactly contracting tensor networks in dimensions larger than two is a formidable task owing to the exponential increase in computational cost as the system size grows. 
In practice, one often resorts to approximate contraction methods, such as the corner transfer matrix renormalization group (CTMRG)~\cite{nishino-corner-1996,nishino-corner-1997,orus-simulation-2009,corboz-simulation-2010,orus-exploring-2012,corboz-competing-2014}, tensor-based coarse graining methods~\cite{levin-tensor-2007,xie-second-2009,gu-tensor-2009,li-linearized-2011,xie-coarse-2012,evenbly-tensor-2015,czarnik-variational-2015,yang-loop-2017,bal-renormalization-2017,hauru-renormalization-2018,chen-exponential-2018}, and boundary MPS methods~\cite{nishino-density-1995,wang-transfer-1997,xiang-thermodynamics-1998,vidal-efficient-2003,orus-infinite-2008,jordan-classical-2008,zauner-stauber-variational-2018,haegeman-unifying-2016,haegeman-diagonalizing-2017,vanderstraeten-tangent-2019}.

Among the approximate contraction methods, the boundary MPS method makes use of the concept of the transfer matrix and the powerful MPS toolkit, in order to approximate the (leading) eigenvector(s) of the transfer matrix as MPS.
When the transfer matrix is Hermitian, because of the existence of the variational principle, the boundary MPS calculation can give very accurate results, which are shown to be consistent with results obtained from other methods~\cite{vanderstraeten-variational-2022}.
In general, one would expect the transfer matrices in tensor networks to be non-Hermitian. 
To deal with the non-Hermitian transfer matrices, a widely employed technique is the biorthogonalization approach~\cite{wang-transfer-1997,xiang-thermodynamics-1998,corboz-simulation-2010,corboz-stripes-2011,huang-biorthonormal-2011-b,huang-biorthonormal-2011,huang-accurate-2012,corboz-competing-2014,fishman-faster-2018}.
The biorthogonalization technique was first introduced in the context of applying density-matrix renormalization group (DMRG) algorithm to non-Hermitian transfer matrices~\cite{wang-transfer-1997,xiang-thermodynamics-1998}, where the biorthogonalization of the left and right eigenvectors was necessary for a proper definition of the reduced density matrix.
This technique has later been applied to the CTMRG contraction of PEPS wavefunctions~\cite{corboz-simulation-2010,corboz-stripes-2011,corboz-competing-2014,fishman-faster-2018}, where a biorthogonalization procedure was required before each CTMRG truncation step. 
However, purely from the point of view of approximating the eigenvector as an MPS, it is not entirely clear why the biorthogonalization technique is necessary for the boundary MPS calculation. 
For example, one may expect that making use of the power method from only one direction in the boundary MPS calculation is adequate for non-Hermitian transfer matrices, as long as the truncation of the boundary MPS itself is accurate enough at each step of power method.
Besides, unlike the single-direction power method, where one can variationally optimize the fidelity during the truncation of the MPS, there is no obvious variational principle available at each truncation step in the biorthogonalization formalism, which makes it difficult to find an optimal truncation scheme. 

In practice, the Hermicity of the transfer matrix is directly affected by the gauge degrees of freedom on the virtual indices.
In the context of PEPS wavefunctions, a gauge fixing condition was proposed recently~\cite{acuaviva-minimal-2022}. 
Another strategy is to impose symmetry constraints on the local tensors of the PEPS ansatz such that the transfer matrix is guaranteed to be Hermitian, at the cost of losing part of the expressibility of the PEPS ansatz~\cite{vanderstraeten-variational-2022}. To assess the importance of such schemes, the impact of the non-Hermicity on the accuracy of the boundary MPS calculation must be better understood.

To address these problems, in this paper, we study the effect of gauge transformations on the boundary-MPS contraction of tensor networks.
In the first part of the paper, we restrict ourselves to the case of local gauge transformations, where the gauge transformation can be represented as local matrices acting on the virtual degrees of freedom.
We show that the effect of the gauge transformation can be understood as an imaginary-time evolution governed by a diagonal Hamiltonian, and the entanglement entropy of the eigenstate tends to be suppressed after the gauge transformation.
Somewhat paradoxically, the suppression in entanglement entropy actually hinders the boundary MPS calculation, as the components of the leading eigenvectors that are relevant to physical observables tend to be suppressed from the entanglement perspective after the gauge transformation, making it more difficult to capture them through the MPS ansatz.
We demonstrate this effect by studying the classical Ising model on the square lattice, where we introduce different local gauge transformations to the tensor network representation of its partition function and study the entanglement properties of the boundary MPSs.
By performing boundary MPS calculations using the transfer matrix after the gauge transformation, we compare the accuracy of the numerical results before and after the gauge transformation.
In the second part of the paper, we go beyond the scope of local gauge transformations and study a more complicated example, namely the antiferromagnetic Ising model on the triangular lattice.
For the partition function of this model, two different tensor network representations~\cite{vanhecke-tangent-2021} exist, which lead to different performances in the boundary MPS calculation. 
We show that these two tensor network representations are connected to each other by a gauge transformation in the form of a matrix product operator.
Similar to cases of local gauge transformations, the distinct numerical performances of the two tensor network constructions can be understood through this gauge transformation, which alters the entanglement properties of the eigenstates, significantly influencing the boundary MPS calculation.

The remainder of the paper is organized as follows. 
In Sec.~\ref{sec:rdm}, we review the concept of the reduced density matrix for non-Hermitian transfer matrices, and point out the importance of the single-sided reduced density matrices on boundary MPS calculations.
In Sec.~\ref{sec:local_gauge}, we analyze the effect of local gauge transformations on the entanglement properties of the eigenstates of the transfer matrix.
In Sec.~\ref{sec:example_square_ising}, we numerically study the effect of local gauge transformations on the boundary MPS calculation using the example of the classical Ising model on the square lattice.
In Sec.~\ref{sec:tiafm}, we study the example of the antiferromagnetic Ising model on the triangular lattice, and demonstrate that the distinct numerical performances of its two tensor network constructions can be understood through a nonlocal gauge transformation.
Section \ref{sec:conclusion} includes a summary and an outlook.

\section{Reduced density matrix} \label{sec:rdm}

In this section, we review the concept of the reduced density matrix for non-Hermitian transfer matrices. 

We start with a transfer matrix $\mathcal{T}$ in a tensor network, and denote its left and right dominant eigenstates as $|\psiL\>$ and $|\psiR\>$.
The expectation value of an operator $O$ defined on the open indices of the transfer matrix can then be evaluated as $\< \psiL |O| \psiR \> / \< \psiL | \psiR \>$.
From this point of view, for the degrees of freedom living on the open indices, we can define the following density matrix
\begin{equation}
    \rho^{\LR} = \frac{|\psiR\> \< \psiL |}{\< \psiL | \psiR \>}.
\end{equation}
To study how the degrees of freedom are correlated with each other, we divide the transfer matrix into two parts, $\calA$ and $\calB$, and introduce reduced density matrix~\cite{brody-biorthogonal-2013}
\begin{equation}
   \rho^{\LR}_{\calA} = \tr_{\calA} \, \rho^{\LR} =  \tr_{\calA} \, \frac{|\psiR\> \< \psiL|}{\< \psiL | \psiR \>}, \label{eq:rdmrhoLR}
\end{equation}
from which one can further obtain the entanglement properties between subsystems $\calA$ and $\calB$, such as entanglement spectrum and the entanglement entropy.

We should note that the definition of the double-sided reduced density matrix above is very subtle.
For a general non-Hermitian transfer matrix, there is no guarantee that the $\rho^\LR$ defined in Eq.~\eqref{eq:rdmrhoLR} is positive-semidefinite, nor is there any guarantee that the eigenvalues of $\rho^\LR$ are non-negative~\cite{cardy-conformal-1985,bianchini-entanglement-2014,bianchini-entanglement-2015,bianchini-entanglement-2016,couvreur-entanglement-2017,dupic-entanglement-2018,tu-renyi-2022}.
In these cases, the definition of the entanglement entropy also needs to be generalized~\cite{tu-renyi-2022}.

Generally speaking, the entanglement properties encoded in $\rho^{\LR}_{\calA}$ differ from those of the eigenstates themselves, which are encoded in the following two single-sided reduced density matrices 
\begin{equation}
   \rho^{\LL} = \tr_{\calA} \frac{|\psiL\> \< \psiL |}{\<\psiL|\psiL\>}, \; 
   \rho^{\RR} = \tr_{\calA} \frac{|\psiR\> \< \psiR |}{\<\psiR|\psiR\>}. 
\end{equation}
The reduced density matrix $\rho_{\LR}$ becomes identical to $\rho_{\LL}$ and $\rho_{\RR}$ when $\mathcal{T}$ is a normal matrix, i.e., $\mathcal{T} \mathcal{T}^\dagger = \mathcal{T}^\dagger \mathcal{T}$, in which case $|\psiL\> = |\psiR\>$.

The reduced density matrices can be obtained by considering the Schmidt decomposition of $|\psiR\>$ and $|\psiL\>$, 
\begin{equation}
    |\psiR\> = \sum_i C^{\rmR}_{ii} |\psiR_{\calA,i} \> |\psiR_{\calB,i}\> , \;
    |\psiL\> = \sum_i C^{\rmL}_{ii} |\psiL_{\calA,i} \> |\psiL_{\calB,i}\> ,
    \label{eq:Schmidt_decomposition}
\end{equation}
where the states defined on the subsystems are orthonormal, $\< \psiR_{\calA, i}| \psiR_{\calA, j} \> = \< \psiR_{\calB, i}| \psiR_{\calB, j} \> = \< \psiL_{\calA, i}| \psiL_{\calA, j} \> = \< \psiL_{\calB, i}| \psiL_{\calB, j} \> = \delta_{i,j}$, 
and $C^\rmR$, $C^\rmL$ are positive-semidefinite diagonal matrices with diagonal entries sorted in descending order. 
The reduced density matrices $\rho_{\RR}$ and $\rho_{\LL}$ can, up to isometries, be simply represented as $(C^\rmR)^\dagger C^\rmR$ and $(C^\rmL)^\dagger C^\rmL$.
The double-sided reduced density matrix $\rho^{\LR}$ can be expressed as
\begin{equation}
    \rho^{\LR}_\calA \doteq M^\calB C^R (M^\calA)^T C^L,
    \label{eq:rhoLR}
\end{equation}
where $M^\calA_{ij} = \< \psiL_{\calA i} | \psiR_{\calA j} \>$, $M^\calB_{ij} = \< \psiL_{\calB i} | \psiR_{\calB j} \>$.
%The two matrices $M^\calA$ and $M^\calB$ will play an important role in our discussion, since they reflect how the different components in $|\psiL\>$ and $|\psiR\>$ are connected.

From the physical point of view, only $\rho^{\LR}_\calA$ properly corresponds to the properties of the system, while the one-sided density matrices $\rho^{\LL}_\calA$ and $\rho^{\RR}_\calA$ are unphysical.
However, from the numerical point of view, these one-sided density matrices become important when we use MPS to approximate the eigenstates, since MPS is an ansatz targeting states with low entanglement and making approximations by truncating components in the state according to their contribution to the entanglement.
Therefore, it is possible to assess the accuracy of the MPS calculation by examining $\rhoLL$, $\rhoRR$, and $\rhoLR$.
More specifically, by analyzing these three reduced density matrices, we can assess how the information relevant for the physical properties of the system is encoded in $|\psiL\>$ and $|\psiR\>$. 
In the most ideal scenario, the transfer matrix is normal, then $\rhoLR$ is identical to $\rhoLL$ and $\rhoRR$, and the usage of MPS ansatz for the eigenstates is optimal in the sense that the entanglement structures in the eigenstates are identical with those encoded in $\rhoLR$.
On the contrary, if the contribution to $\rhoLR$ mainly comes from the components only making minor contribution to the entanglement in $|\psiL\>$ and $|\psiR\>$, they are likely to be discarded in the MPS approximation, and the MPS calculation may become inaccurate or even problematic. 

\section{Effect of gauge transformation: local gauges} \label{sec:local_gauge}

In this section, we analyze the effect of local gauge transformations on the entanglement properties of the eigenstates of the transfer matrix. 

We start our discussion with a MPO $\mathcal{T}_0$ which is a normal matrix, whose dominant eigenvector is denoted as $|\psiL_0\> = |\psiR_0\> = |\psi_0\>$. 
If we introduce a reversible gauge transformation $\mathcal{P}$ to the MPO local tensor, the resulting MPO $\mathcal{T}$ can be written as
\begin{equation}
    \mathcal{T} = \mathcal{P} \mathcal{T}_0 \mathcal{P}^{-1}.
\end{equation}
Here, we restrict ourselves to local gauge transformations, where $\mathcal{P}$ can be represented as $\mathcal{P} = P^{\otimes N}$, where $P$ is the local matrix acting on the open indices of the MPO, and $N$ is the length of the MPO. 
If $P$ is unitary, or if $[\mathcal{P}, \mathcal{T}_0] = 0$, then $\mathcal{T}$ remains to be a normal matrix. 
In general, $\mathcal{T}$ is no longer normal, and the left and right eigenvectors become 
\begin{equation}
    |\psi_R\> = \mathcal{P} |\psi_0\>, \quad |\psi_L\> = (\mathcal{P}^{-1})^\dagger |\psi_0\>.\label{eq:fixed_point_after_gauge}
\end{equation}
The local gauge transformation $\mathcal{P}$ only introduces a similarity transformation to $\rho^{\LR}_\calA$, which does not change its eigenvalues. 
In contrast, the entanglement properties of the eigenstates $|\psiL\>$ and $|\psiR\>$ are altered by the gauge transformation.
The effect of the gauge transformation can be understood by considering the singular value decomposition of the matrix $P= U S V^\dagger$. 
As $U$ and $V$ are unitary matrices, only the diagonal part $S$ will alter the entanglement structure of $|\psiR\>$ and $|\psiL\>$. 
Since $S$ is positive definite, we can always write $S$ as $S = \exp(-\Lambda)$, where $\Lambda$ is a real diagonal matrix.
From this point of view, the gauge transformation $\mathcal{P}$ can be regarded as an imaginary-time evolution governed by the diagonal Hamiltonian $\sum_{j=1}^N \Lambda_j$ up to local unitaries.
According to Eq.~\eqref{eq:fixed_point_after_gauge}, this imaginary-time evolution will drive the eigenstates towards two orthogonal product states, which are the ground state and the highest-energy state of the diagonal Hamiltonian, respectively.
The strength of the time evolution can be characterized by the gap $\Delta$ in the diagonal Hamiltonian.
As the gap $\Delta$ increases, the entanglement properties of the eigenstates undergo more pronounced changes. 
When $\Delta \gg 1$, the eigenstates $|\psiL\>$ and $|\psiR\>$ become very close to two orthogonal product states.

When the gauge transformation suppresses the entanglement entropy in the eigenvector, one may anticipate that the boundary MPS calculation becomes easier, as MPS is a more suitable ansatz for low-entanglement states.
However, our calculation aims not only to approximate the eigenstate but, more importantly, to compute the properties of the system using these eigenstates.
From this point of view, the gauge transformation actually has a negative effect on the MPS calculation, because the components in $|\psiL\>$ and $|\psiR\>$ that are of physical relevance, i.e., those that contribute to $\rhoLR$, are suppressed by the gauge transformation.
Roughly speaking, after the gauge transformation, the SVD decomposition of the eigenstates [recall Eq.~\eqref{eq:Schmidt_decomposition}] can be expressed as
\begin{align}
|\psiR\> &= \sum_{i} C_i^{\rmR,>} |\psi_i^{\rmR,>}\> + \sum_{j} C_j^{\rmR,<} |\psi_j^{\rmR,<}\>, \label{eq:psiR_ill_conditioned}\\ 
|\psiL\> &= \sum_{i} C_i^{\rmL,>} |\psi_i^{\rmL,>}\> + \sum_{j} C_j^{\rmL,<} |\psi_j^{\rmL,<}\>. \label{eq:psiL_ill_conditioned}
\end{align}
The components $|\psi^{\rmR,>}\>$ and $|\psi^{\rmL,>}\>$ are introduced by the gauge transformation, which satisfies $\<\psi_i^{\rmL, >}|\psi_{i'}^{\rmR, >}\> = 0$, while the rest of the components $|\psi^{\rmR,<}\>$ and $|\psi^{\rmL,<}\>$ are the ones that are of physical relevance. 
With the strength of the gauge transformation (the gap $\Delta$ of the diagonal Hamiltonian) increases, the coefficients $C_j^{\rmR,<}$ and $C_j^{\rmL,<}$ decays exponentially, and the non-physical components $|\psi_i^{\rmR, >}\>$ and $|\psi_i^{\rmL, >}\>$ gradually become dominant in the eigenstates.
The suppression of $|\psi_i^{\rmR, <}\>$ and $|\psi_i^{\rmL, <}\>$ clearly makes it more difficult to calculate the physical properties of the system in boundary MPS calculations. 

\section{Example: classical Ising model on square lattice} \label{sec:example_square_ising}

In this section, using the example of the two-dimensional ferromagnetic square-lattice classical Ising model at the critical temperature, we numerically study the effect of a local gauge transformation in the virtual degrees of freedom.

\subsection{Setup of the calculation}
In the standard construction of the tensor network representation of the partition function, the transfer matrix $\mathcal{T}_0$ is a Hermitian matrix with identical left and right eigenvectors, denoted as $|\psi_0\>$.
We then introduce a local gauge transformation $\mathcal{P} = P^{\otimes N}$, where $P$ represents the local matrix acting on the virtual degrees of freedom, and $N$ is the length of the transfer matrix~\cite{fishman-faster-2018}.
This transformation modifies the transfer matrix to $\mathcal{T} = \mathcal{P} \calT_0 \mathcal{P}^{-1}$ and transforms the left and right eigenvectors into $|\psiR\> = \mathcal{P} |\psi_0\>$ and $|\psiL\> = \mathcal{P}^{-1} |\psi_0\>$, respectively. 
Since the original transfer matrix is Hermitian, we can efficiently obtain the MPS approximations for the original eigenvector $|\psi_0\>$  using the VUMPS algorithm, from which we can directly obtain the MPS approximations for the left and right eigenvectors $|\psiL\>$ and $|\psiR\>$ after the gauge transformation.
This allows us to accurately calculate the corresponding single-sided reduced density matrices, and calculate the entanglement properties of the eigenstates before and after the gauge transformation.

In the following, we will first calculate and compare the entanglement properties of the eigenstates after different gauge transformations.
Secondly, we will perform boundary MPS calculations directly using the gauge-transformed transfer matrix $\mathcal{T}$, without resorting to the original eigenvector.  
We will compare the accuracy of the numerical results and interpret the results using the entanglement properties of the eigenstates.

\subsection{Entanglement in the eigenvectors} 
% how to obtain the eigenvectors before and after the gauge transformation
% what do you expect for the entanglement structure of the eigenvectors before the gauge transformation
% what happens to the entanglement structure of the eigenvectors after the gauge transformation
% why is the entanglement structure related to the performance of boundary MPS calculation
Before the gauge transformation, since the transfer matrix is Hermitian, the entanglement properties of the eigenstates are well known.
Since the system is critical, the entanglement entropy of the fixed-point MPS is expected to grow logarithmically with the correlation length that is induced by the finite bond dimension~\cite{pollmann-theory-2009}.  

%As discussed in Sec.~\ref{sec:local_gauge}, a local gauge transformation can, up to local unitaries, be understood as an imaginary-time evolution governed by a diagonal Hamiltonian, and the entanglement entropy of the boundary MPS is expected to suppressed by the gauge transformation. 
%In the following, we will numerically verify this expectation.
Next, we examine the entanglement properties of the eigenstates after the local gauge transformation.
As discussed in Sec.~\ref{sec:local_gauge}, it suffices to consider local matrices acting on the virtual degrees of freedom as $P = \exp(-\tau Q)$, where $Q$ is a Hermitian matrix.
In our calculation, we will choose different values of $\tau$ and two different choices for $Q$: $Q=\sigma^x$ and $M=\sigma^z$, with $\sigma^x$ and $\sigma^z$ being the Pauli matrices.

\begin{figure}[!htb]
    \centering
    \resizebox{\columnwidth}{!}{\includegraphics{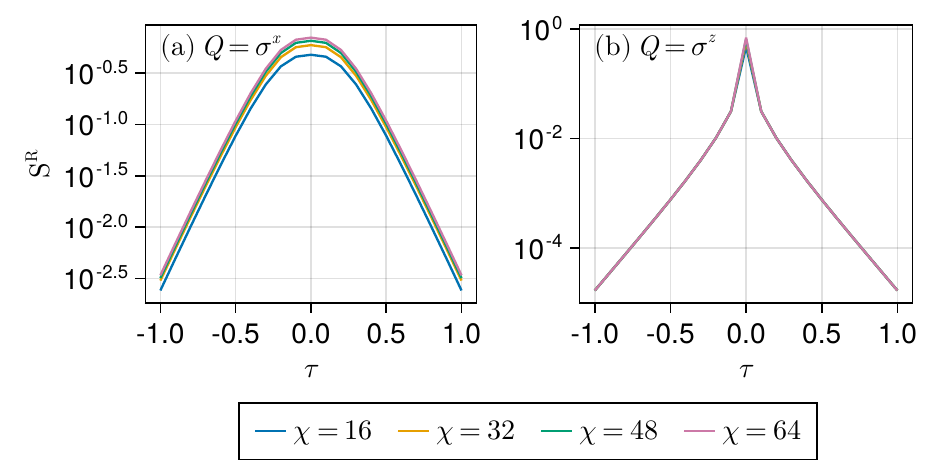}}
    \caption{The entanglement entropy $S^{\text{R}}$ in the right boundary MPS after the gauge transformation with $P=\exp(-\tau M)$ where (a) $M = \sigma^x$ and (b) $M = \sigma^z$. Note that $S^{\text{L}}(\tau)=S^{\text{R}}(-\tau)$. 
    In our calculation, different bond dimensions of the boundary MPS are used, which are marked by different colors. 
    Note that the results for different $\chi$ in (b) almost overlap with each other.}
    \label{fig:EE-local-gauge}
\end{figure}

\begin{figure}[!htb]
    \centering
    \resizebox{\columnwidth}{!}{\includegraphics{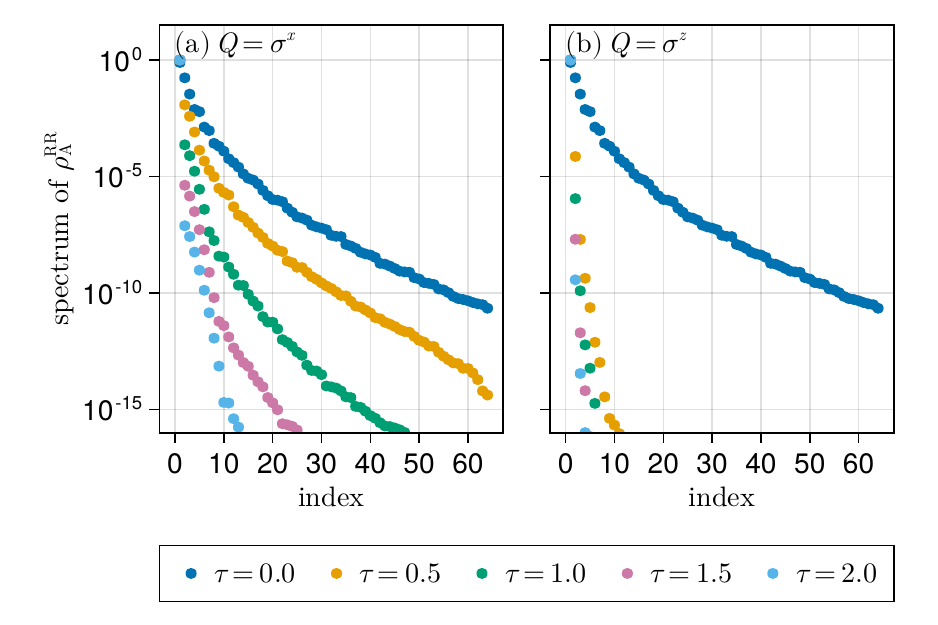}}
    \caption{The spectrum of the reduced density matrix of the right eigenvector after the gauge transformation with (a) $ M = \sigma^x$ and (b) $M = \sigma^z$. 
    The bond dimension of the boundary MPS is $\chi=64$.
    The entries with value smaller than $10^{-16}$ are not shown.
    }
    \label{fig:ES-local-gauge}
\end{figure}

In Figs.~\ref{fig:EE-local-gauge} and \ref{fig:ES-local-gauge}, we show the entanglement entropy and the spectrum of the reduced density matrix for the right eigenvector after applying the gauge transformations. Note that the right eigenvector for a gauge transformation with parameter $\tau$ is identical to the left eigenvector for a gauge transformation with parameter $-\tau$. From Fig.~\ref{fig:EE-local-gauge}, we observe that the local gauge transformation suppresses the entanglement entropy of the boundary MPS. 
Except for small values of $|\tau|$, the entanglement entropy decays exponentially with $|\tau|$. 
Results for the spectrum of the reduced density matrix in Fig.~\ref{fig:ES-local-gauge} further illustrate the effect of the gauge transformation: with the strength of the entanglement entropy increasing, the eigenstate becomes closer to a product state.
These results are consistent with the analysis in Sec.~\ref{sec:local_gauge}.
We note that a similar effect is also observed in real-time dynamical simulations, where nonunitary local gauge transformations are used to suppress the growth of entanglement in the MPS~\cite{PhysRevB.106.104306}. 

We also observe that the results depend on the choice of the local matrix $Q$.
In the case of $Q = \sigma^z$, the entanglement entropy of the boundary MPS clearly decays faster with $\tau$ than the case of $Q = \sigma^x$.

\begin{figure}[!htb]
    \centering
    \resizebox{\columnwidth}{!}{\includegraphics{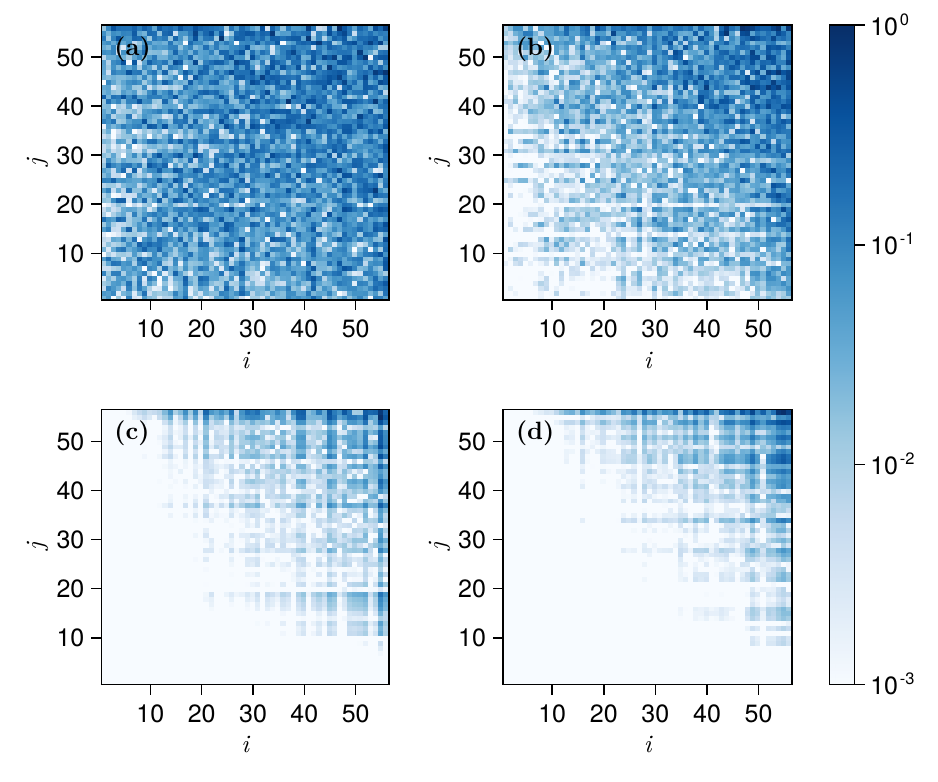}}
    \caption{The absolute value of the matrix element $M_{ij}^A$ after the gauge transformation with (a) $\tau = 0.5, Q = \sigma^x$, (b) $\tau = 1.0, Q = \sigma^x$, (c) $\tau = 0.5, Q = \sigma^z$, and (d) $\tau = 1.0, Q = \sigma^z$.
    The bond dimension of the boundary MPS is $\chi=64$. 
    In each plot, the matrix $M^A$ is normalized such that $\max_{i,j} |M_{ij}^A| = 1$, and elements with a norm smaller than $10^{-3}$ are not shown. 
    }
    \label{fig:scattering-mat}
\end{figure}

In Sec.~\ref{sec:local_gauge}, we also argued that, after the gauge transformation, the components in $|\psiL\>$ and $|\psiR\>$ that are of physical relevance will be suppressed. 
We can visualize this effect by examining the matrix elements of the matrices introduced in Eq.~\eqref{eq:rhoLR}.
From Eq.~\eqref{eq:rhoLR}, it is clear that the matrices $M^A$ and $M^B$ determine the contributions of the different Schmidt vectors in $|\psiL\>$ and $|\psiR\>$ to the reduced density matrix $\rhoLR$.
In the MPS formalism, these $M$ matrices can be easily calculated by converting the boundary MPSs to the mixed-canonical form.
Here, we only check the norms of the matrix elements of $M^A$. 
In Fig.~\ref{fig:scattering-mat}, we show the norm of the matrix elements of $M^A$ after the gauge transformation with different values of $\tau$ and $Q$.
By comparing Fig.~\ref{fig:scattering-mat} with Figs.~\ref{fig:EE-local-gauge} and \ref{fig:ES-local-gauge}, we observe that as the entanglement of the boundary MPS is more effectively suppressed, the finite-valued matrix elements in $M^A$ tend to concentrate in the regions corresponding to larger values of the matrix indices $i$ and $j$. 
These regions correspond to the components in $|\psiL\>$ and $|\psiR\>$ that correspond to the tail of the entanglement spectrum. 

\subsection{Numerical results}

Next, we compare the accuracy of the numerical calculation for the transfer matrix after the gauge transformation.
In the numerical calculation, as there is no variational principle, we separately obtain the MPS approximations for the left and right eigenstates $|\psiL\>$ and $|\psiR\>$ using the power method. 
At each power method step, we compress the MPS to a certain bond dimension using two different strategies, namely the infinite time-evolving block decimation (iTEBD)~\cite{orus-infinite-2008,mcculloch-infinite-2008} and the variational optimization of matrix product states using tangent space methods (VOMPS)~\cite{vanhecke-tangent-2021}. 
In the iTEBD calculation, the number of the power method steps is 500. 
At each step, we convert the MPS to the canonical form and truncate it by only retaining Schmidt coefficients greater than $10^{-8}$. 
The bond dimension of the boundary MPS thus depends on its entanglement spectrum and will vary during the power method calculation.
In the VOMPS calculation, we incrementally increase the bond dimension of the boundary MPS, performing calculations for $\chi = 2, 4, 8, 16, 32, 64$ in sequence.
For each bond dimension, we conduct 250 power method steps and, at each step, truncate the MPS by variational optimization.
The result obtained with the smaller bond dimension can be used to initialize the calculation for the larger bond dimension.

To assess the quality of the numerical results, we compute the relative error of the free energy.
Since there is no variational principle in $\<\psiL|\calT|\psiR\>/\<\psiL|\psiR\>$, the free energy obtained in this way may have error cancellation and the corresponding relative error obtained cannot be used as a reliable indicator of the accuracy of the calculation.
Instead, we only make use of $|\psiR\>$ and compute the free energy density using the original transfer matrix 
\begin{equation}
    f = - \frac{1}{L\beta} \ln\frac{\<\bar{\psi}|\calT_0|\bar{\psi}\>}{\<\bar{\psi}|\bar{\psi}\>}, \; |\bar{\psi}\> = \calP^{-1}|\psiR\>,
    \label{eq:free_energy_T0} 
\end{equation}
where $L$ is the length of the MPO.
Since $\calT_0$ is a normal matrix, the relative error of the free energy obtained using Eq.~\eqref{eq:free_energy_T0} can indicate the accuracy of the boundary MPS $|\psiR\>$.
A similar calculation can be done for $|\psiL\>$.

Additionally, we calculate the quantity
\begin{equation}
    V(\mathcal{T}, |\psiR\>) = \frac{1}{L} \ln \left[
        \frac{\<\psiR| \mathcal{T}^\dagger \mathcal{T} |\psiR\> \<\psiR| \psiR \>}{\<\psiR| \mathcal{T} |\psiR\> \<\psiR| \mathcal{T}^\dagger |\psiR\>}
    \right].
    \label{eq:variance} 
\end{equation}
which quantifies how closely the state $|\psiR\>$ aligns with a right eigenvector of the transfer matrix. The argument of the $\ln$ is lower bounded by $1$ using Cauchy-Schwarz, or thus, $V(\mathcal{T}, |\psiR\>)\geq 0$ with equality only if $|\psi\>$  is an eigenvector of $\mathcal{T}$.
The whole expression can be interpreted as size-consistent generalisation of the variance, in the sense that for $\mathcal{T} = 1 + \epsilon \mathcal{H}$, the argument of the logarithm would reduce to $1 + \epsilon^2 \lVert(\mathcal{H}-\<\mathcal{H}\>) |\psiR\>\rVert^2$. A similar definition can be given for the left boundary MPS.
Although the variance cannot distinguish between a state that is close to the dominant eigenvector and one that is close to an ordinary eigenvector, in our power method calculation, it can still be useful in assessing the quality of the MPS approximation. 
The variance can be computed efficiently using the MPS representation of the state $|\psi\>$ and the MPO representation of the transfer matrix $\calT$.

\begin{figure}[!htb]
    \centering
    \resizebox{\columnwidth}{!}{\includegraphics{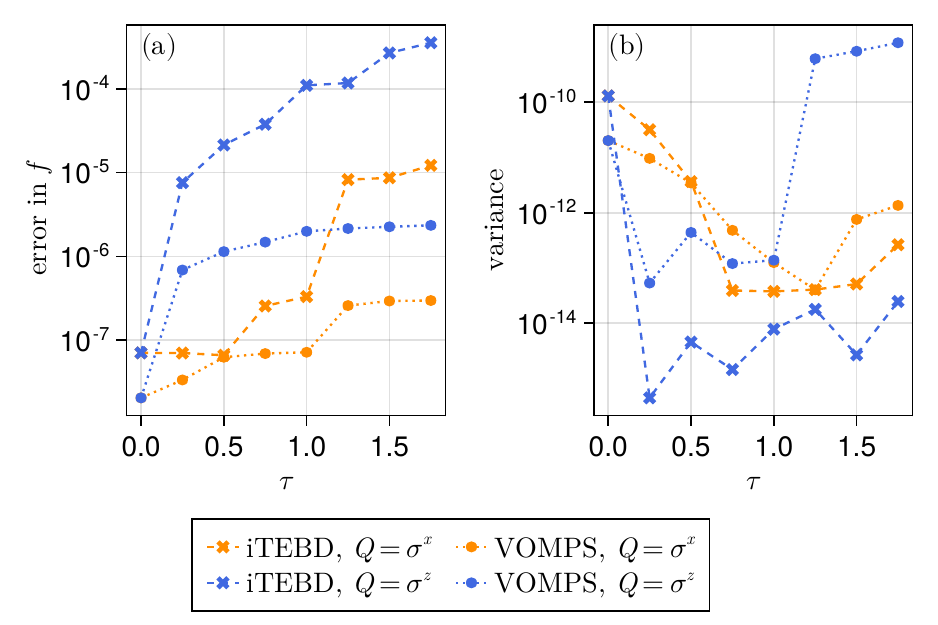}}
    \caption{The results for (a) relative error in the free energy and (b) the variance of the right boundary MPS.
    The gauge transformations with different choices of $Q$ and values of $\tau$ are employed. 
    The results obtained using iTEBD and VOMPS are shown with different symbols.
    }
    \label{fig:results-local-gauge}
\end{figure}

We show the results of the numerical calculation in Fig.~\ref{fig:results-local-gauge}.
The numerical results for the relative error in the free energy align well with our previous analysis [see Fig.~\ref{fig:results-local-gauge} (a)].
With $\tau$ increasing, the relative error in the free energy increases. %, indicating that the accuracy of the boundary MPS deteriorates.
This can be understood from our previous discussion: In each iTEBD step, the MPS is truncated according to the entanglement spectrum of the boundary MPS, which, when $\tau$ is large, is actually dominated by the components of no physical relevance.
One can also observe that the relative error in the case of $Q=\sigma^z$ increases more rapidly than the case of $Q=\sigma^x$, which can be attributed to the fact that the entanglement entropy of the boundary MPS is more effectively suppressed in the case of $Q=\sigma^z$. 
On the other hand, the VOMPS calculation uses a different strategy, where the boundary MPS is truncated by variational optimization at each power method step.
Therefore, the VOMPS simulation is less sensitive to the gauge transformation, although its accuracy also deteriorates with $\tau$ increasing.
During the variational optimization at each power method iteration, we optimize the fidelity between the truncated MPS and the original one. 
Recall Eqs.~\eqref{eq:psiR_ill_conditioned} and \eqref{eq:psiL_ill_conditioned}, when $\tau$ increases, the components of physical relevance make smaller contributions to the fidelity, and a small change in the fidelity may lead to a significant deviation in these physical components.

\begin{figure}[!htb]
    \centering
    \resizebox{\columnwidth}{!}{\includegraphics{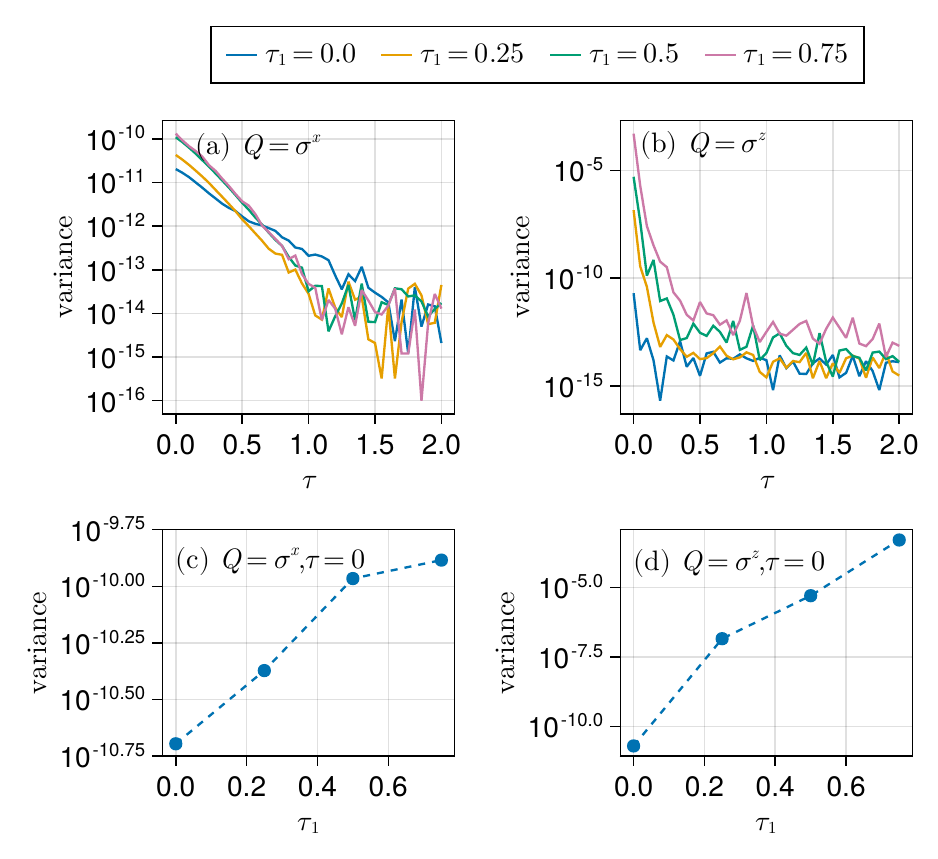}}
    \caption{The results for $V(\calT_\tau, |\psiR_1\>)$ with different values of $\tau$ and $\tau_1$. 
    }
    \label{fig:variances-change}
\end{figure}

Unlike the relative error in the free energy, when increasing $\tau$, the variance of the boundary MPS decreases with $\tau$ (for the data with $\tau < 1.5$) [see Fig.~\ref{fig:results-local-gauge} (b)], indicating the boundary MPS that we obtained becomes closer to an eigenvector of the transfer matrix.
This is actually not surprising since the entanglement entropy of the eigenstate decreases with $\tau$, which makes it easier to be approximated by an MPS.
On the other hand, in this case, obtaining a ``better'' approximation of the eigenstate does not lead to a more accurate calculation of the physical properties of the system, as the components of physical relevance are already suppressed by the gauge transformation.
To further illustrate this point, at each $\tau$, we calculate the variance of the boundary MPS obtained by applying the corresponding gauge transformation on the fixed-point MPS at another value of $\tau$. 
More specifically, we calculate the quantity $V(\calT_\tau, |\psiR_1\>)$ [see Eq.~\eqref{eq:variance}], where $\calT_\tau = \calP_\tau \calT_0 \calP_\tau^{-1}$, and $|\psiR_1\> = \calP_{\tau-\tau_1} |\psiR_{\tau_1}\>$, with $|\psiR_{\tau_1}\>$ denoting the fixed-point MPS obtained with power method using the MPO $\calT_{\tau_1} = \calP_{\tau_1} \calT_0 \calP_{\tau_1}^{-1}$. 
The results are shown in Fig.~\ref{fig:variances-change}.
From Fig.~\ref{fig:variances-change}, we see that the relative error in the free energy is only in accordance with the variance of the boundary MPS when $\tau$ is close to 0, i.e., when $\calT_{\tau}$ is close to a normal matrix. 
When $\tau$ becomes large, the variances of different MPSs all decrease with $\tau$, as their difference between each other are suppressed by the gauge transformation, and they all become very close to the dominant eigenvector the gauged transfer matrix $\calT_\tau$.
Furthermore, Fig.~\ref{fig:variances-change} also indicates that a lower variance no longer imply a better physical result when the transfer matrix is far away from a normal matrix.  

\subsection{Discussion} \label{sec:local-gauge-discussion}

In practice, the local gauge degrees of freedom can easily be fixed before the boundary MPS calculation. 
We consider the following measure of the normality of the transfer matrix
\begin{equation}
    N(\calT) = F\left(\calT \calT^\dagger , \calT^\dagger \calT\right), 
    \label{eq:normality}
\end{equation}
where $F(\calT_1, \calT_2)$ represents the fidelity between two matrices $\calT_1$ and $\calT_2$, which is defined as
\begin{equation}
    F(\calT_1, \calT_2) = \frac{\tr\left(\calT_1^\dagger \calT_2\right)}{\sqrt{\tr\left(\calT_1^\dagger \calT_1\right)\tr\left(\calT_2^\dagger \calT_2\right)}}.
    \label{eq:fidelity_matrices}
\end{equation}
The normality $N(\calT)$ ranges from 0 and 1, with 1 indicating perfect normality.

Suppose there exists a local gauge transformation $\calP$ such that $\tilde{\calT} = \calP \calT \calP^{-1}$ is a normal matrix, then we can obtain this $\calP$ by maximizing the normality measure $N(\tilde{\calT})$.
In practice, it is not easy to directly parametrize the local matrix $P$ in $\calP$. 
Notice that requiring $\tilde{\calT}^\dagger \tilde{\calT} = \tilde{\calT} \tilde{\calT}^\dagger$ is equivalent to requiring 
$\calG \calT \calG^{-1} \calT^\dagger \calG = \calT^\dagger \calG \calT$, where $\calG = \calP^\dagger \calP$.
Then we can obtain the optimal $\calP$ by maximizing $F(\calG \calT \calG^{-1} \calT^\dagger \calG , \calT^\dagger \calG \calT)$, during which we can parametrize the local matrix in $\calG$ as an exponential of a Hermitian matrix. 

Alternatively, we note that, recently, a new method for eliminating local gauge degrees of freedom has emerged --- the minimal canonical form~\cite{acuaviva-minimal-2022}.
In the minimal canonical form, the gauge degrees of freedom are removed by selecting a gauge in which the Frobenius norm of the local tensor is minimized.
This gauge condition is applicable to PEPS in two or higher dimensions as well as tensor networks without physical indices, such as those representing partition functions or PEPS overlaps. While we numerically found that imposing this minimal canonical form on the MPO tensor of gauge-transformed square lattice Ising model undoes the gauge transform and restores the original tensor, which yields a normal MPO, we were not able to prove that this result holds more generally, except for the case of a single-site system, i.e. a matrix \footnote{For a matrix $M$, consider $M_\epsilon$ = $\mathe^{-\epsilon Q} M \mathe^{\epsilon Q}$, where $\epsilon \ll 1$ and $Q$ is a Hermitian matrix. 
Then $\tr(M_\epsilon^\dagger M_\epsilon) = \tr(M^\dagger M) + 2\epsilon \tr(Q [M^\dagger, M]) + 2 \epsilon^2 \tr([M, Q]^\dagger [M, Q])$.
If the matrix $M$ is normal, then $[M^\dagger, M] = 0$ and the first order term vanishes. As the second order term is positive, this indicates that the point where $M$ is normal is a minimum.}.

We emphasize that the methods discussed above can only handle local gauge transformations.
In general, when the transfer matrix is not normal, it may require a nonlocal gauge transformation to transform it to a normal matrix. 
In those cases, one can still use the methods discussed above to make the transfer matrix more ``normal''. 
However, it cannot be expected to necessarily improve the computation accuracy.

\begin{figure}[!htb]
    \centering
    \resizebox{0.667\columnwidth}{!}{\includegraphics{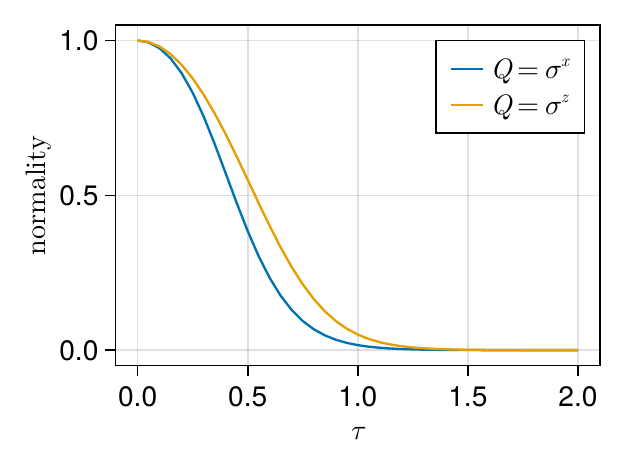}}
    \caption{The normality measures of the transfer matrix $\calT$ for different $\tau$ and $Q$.}
    \label{fig:normality-change}
\end{figure}

Furthermore, we should point out that, in general, the normality measure \eqref{eq:normality} cannot be used to assess the ``quality'' of the MPO unless it is equal or very close to 1.
As a straightforward example, we show the normality measure of the transfer matrix of our current example in Fig.~\ref{fig:normality-change}.
According to Fig.~\ref{fig:normality-change}, for the same value of $\tau$, the normality measure for the transfer matrix with $Q=\sigma^x$ is slightly lower than that with $Q=\sigma^z$. 
However, from our previous discussion, the numerical performance in the case of $Q=\sigma^z$ is much worse than that of $\sigma^x$.

\section{Example: Antiferromagnetic classical Ising model on triangular lattice} \label{sec:tiafm}

In this section, we look at a more complicated example --- the antiferromagnetic classical Ising model on the triangular lattice~\cite{wannier-antiferromagnetism-1950}, where nonlocal gauge transformations are involved.
In this model, one can construct two different tensor network representations for the partition function, which are equivalent from the physical point of view, but have different numerical performance~\cite{vanhecke-solving-2021}.
In the following, we will first briefly review the two different tensor network representations, and then demonstrate that these two representations can be connected by a nonlocal gauge transformation.
Finally, we will analyze the entanglement properties of the eigenstates and discuss the effect of the nonlocal gauge transformation on the boundary MPS calculation.

\subsection{Tensor network representations} 

Because of the frustration in the system, the ground state is highly degenerate. 
Each ground-state configuration should satisfy the so-called ice rule, i.e., on each triangle, there should be two spins pointing up and one spin pointing down, or vice versa, and no ferromagnetic triangles are allowed. 
At zero temperature, the calculation of the partition function is reduced to a counting problem, namely, counting the number of spin configurations satisfying the ice rule.
Along this line, by encoding the ice rules into the local tensors, one can construct the tensor network representation for the partition function. 

\begin{figure}[!htb]
    \centering
    \resizebox{\columnwidth}{!}{\includegraphics{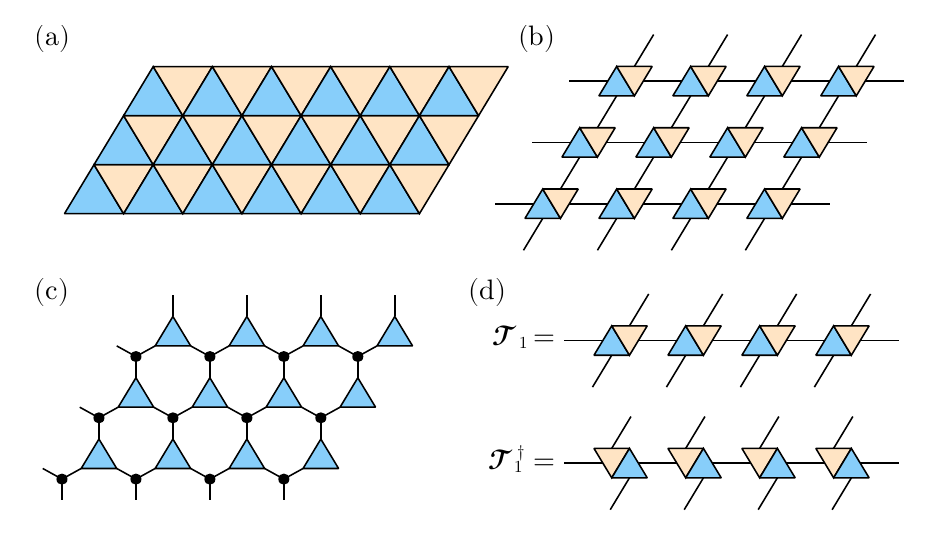}}
    \caption{(a) The triangular lattice and the ice rules. 
    (b) The first tensor network representation of the partition function at zero temperature, which encodes the ice rules on both the upward and downward triangles. 
    (c) The second tensor network representation of the partition function, which encodes the ice rules only on the upward triangles. 
    (d) The transfer matrix $\calT_1$ in the first construction and its Hermitian conjugate $\calT_1^\dagger$.}
    \label{fig:tensor-constructions-TLAIM}
\end{figure}

As pointed out by Ref.~\cite{vanhecke-solving-2021}, there exist two different ways of encoding the ice rules into the local tensors [see Fig.~\ref{fig:tensor-constructions-TLAIM} (a)-(c)].
In the first construction, the ice rules on both upward and downward triangles are explicitly encoded into the local tensors.
As shown in Fig.~\ref{fig:tensor-constructions-TLAIM} (b), each rank-4 tensor corresponds to an upward triangle and a downward triangle, and each index of the local tensor corresponds to two neighboring spins. 
In the second construction, only the ice rules on the upward triangles are encoded into the local tensors. 
These ice rules are encoded in the rank-3 tensors [represented by the blue triangles in Fig.~\ref{fig:tensor-constructions-TLAIM} (c)], where each index corresponds to one spin. 
These ice-rule tensors are connected to each other by the $\delta$-tensors [represented by black dots in Fig.~\ref{fig:tensor-constructions-TLAIM} (c)].  
While the ice rule for the downward triangle is not explicitly incorporated into the construction of the local tensor, it is, in fact, implicitly accounted for~\cite{vanhecke-solving-2021}. 
Requiring ice rule adherence on all upward triangles also implicitly imposes constraints that compel compliance with the ice rule for all downward triangles.
Although these two tensor network constructions describe the same physical system and should have physical equivalence, numerical simulations reveal entirely different performance: VUMPS calculations converge only when using the first construction method, while it fails for the second one~\cite{vanhecke-solving-2021}.

\begin{figure}[!htb]
    \centering
    \resizebox{\columnwidth}{!}{\includegraphics{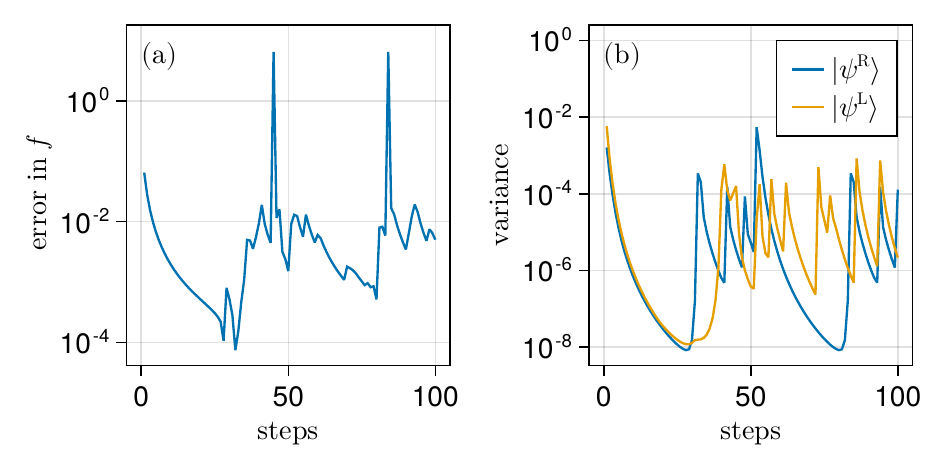}}
    \caption{Results of applying the power method to the transfer matrix $\calT_2$. At each step, the MPS is truncated using the VOMPS method. The bond dimension of the boundary MPS is $\chi=32$.
    At each power method step, we calculate: 
    (a) the relative error in the free energy, which is obtained using $\<\psiL|\calT_2|\psiR\>/\<\psiL|\psiR\>$; and
    (b) the variance of the boundary MPSs [c.f. Eq.~\eqref{eq:variance}].
    }
    \label{fig:vomps-frustrated-mpo}
\end{figure}

From Fig.~\ref{fig:tensor-constructions-TLAIM} (d), we can see that the transfer matrix $\calT_1$ in the first construction satisfies $\calT_1^\dagger = \calT_1 \calU$, where $\calU$ is the translation operator. 
Since $\calT_1$ is translational invariant, i.e., $[\calT_1, \calU] = 0$, it then becomes clear that $\calT_1^\dagger \calT_1 = \calT_1 \calT_1^\dagger$.
Hence, $\calT_1$ is a normal matrix, and the success of the VUMPS calculation thus becomes very natural from this point of view.
On the other hand, it is easy to verify that the transfer matrix $\calT_2$ in the second construction is not a normal matrix.
However, the lack of normality of $\calT_2$ cannot fully explain the failure of the boundary MPS calculation.
Actually, as shown in Fig.~\ref{fig:vomps-frustrated-mpo}, even the power method cannot converge for $\calT_2$.

\subsection{Analysis of transfer matrices}

To further understand the difference between the two transfer matrices, we consider how they act on a single row of spin configurations.

Before our discussion, we notice that the local physical spaces for $\calT_1$ and $\calT_2$ are different. 
For $\calT_1$, each index of the local tensor corresponds to two neighboring spins, while for $\calT_2$, each index corresponds to one spin.
This can easily be fixed by introducing a transformation to the indices of $\calT_1$, which maps $\{(s_1, s_2), (s_2, s_3), \ldots\}$ to $\{s_1, s_2, s_3, \ldots\}$.
Such transformation can be written as an MPO, and by applying it to $\calT_1$, one can transform the MPO $\calT_1$ to a new MPO, which has the same local physical space as $\calT_2$.
Since the properties of $\calT_1$ are unchanged by this ``trivial'' transformation, in the following, we will call the transformed MPO as $\calT_1$ for simplicity. 

\begin{figure}[!htb]
    \centering
    \resizebox{\columnwidth}{!}{\includegraphics{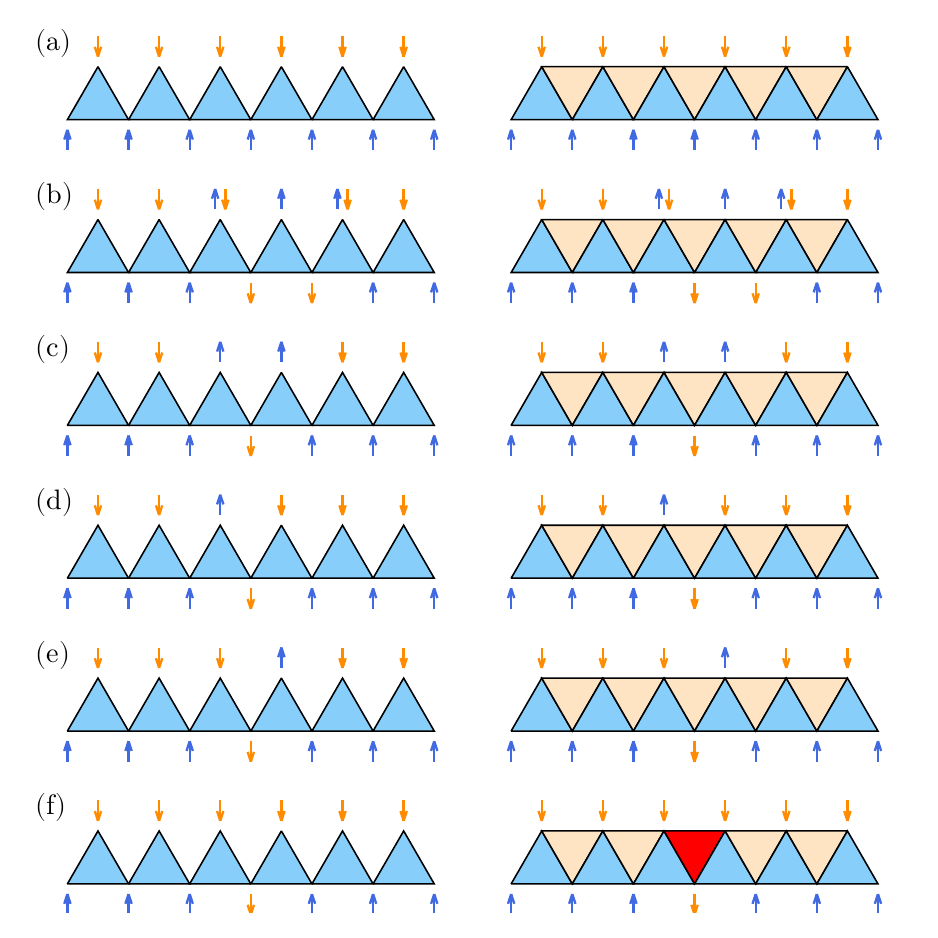}}
    \caption{Analysis of the transfer matrices $\calT_2$ (left) and $\calT_1$ (right). In each figure, the transfer matrix acts on the bottom row of spins, and the top row of spins is the result of the action of the transfer matrix. 
    In the analysis, the length of the transfer matrix is $L=6$ and the periodic boundary condition is employed.
    (a) The transfer matrix acting on a row of spins with no domain walls.
    (b) The transfer matrix acting on a row of spins with two separate domain walls (the distance between two domain walls is larger than 1).
    The symbol $\uparrow \downarrow$ represents that the outcome spin on that site can be either spin-up or spin-down. 
    (c-f) The transfer matrix acting on a row of spins with two neighboring domain walls. 
    The only difference between $\calT_1$ and $\calT_2$ is shown in (f): here, the outcome of $\calT_2$ is forbidden in $\calT_1$, as it violates the ice rule on one of the downward triangles, which is marked by the red color.
    }
    \label{fig:transfer-matrix-analysis}
\end{figure}

By applying the ice rules on the local triangles, we can subsequently analyze the action of the transfer matrices $\calT_1$ and $\calT_2$ on different spin configurations: on spin configuration without domain walls [in Fig.~\ref{fig:transfer-matrix-analysis} (a)], on spin configurations with two separate domain walls [in Fig.~\ref{fig:transfer-matrix-analysis} (b)], and on spin configurations with two neighboring domain walls [in Fig.~\ref{fig:transfer-matrix-analysis} (c-f)]. 
From Fig.~\ref{fig:transfer-matrix-analysis}, it is clear that the effects of $\calT_1$ and $\calT_2$ are identical in most cases: In Fig.~\ref{fig:transfer-matrix-analysis} (a-e), the two transfer matrices give identical results. 
The only difference between the two transfer matrices is shown in (f): the transfer matrix $\calT_2$ allows the elimination of two neighboring domain walls, while $\calT_1$ does not.
This additional domain-wall-elimination process in $\calT_2$ is clearly unphysical, as it violates the ice rule on the downward triangles.
We note that a similar analysis has been conducted for the triangular lattice antiferromagnetic Ising model on an infinite cylindrical geometry~\cite{nourhani-communicating-2018}.

From the analysis above, it also becomes clear that the number of domain walls is a $\Uone$ charge of the system~\cite{blote-antiferromagnetic-1993,jiang-ordering-2006,smerald-spinliquid-2018}. 
The first transfer matrix $\calT_1$ never changes the number of domain walls in the spin configuration, while $\calT_2 = \calT_1 + \Delta \calT$ breaks the $\Uone$ symmetry since it contains additional terms $\Delta \calT$ which can eliminate two neighboring domain walls. 
However, we note that these additional terms $\Delta \calT$ are ``unidirectional'', i.e., they only allow the elimination of two neighboring domain walls, but not the creation of any domain walls. 
Recall the partition function $Z=\sum_{\alpha} \<\alpha|\calT^N | \alpha\>$ ($N$ being the number of rows in the lattice, and the summation iterate over all possible spin configurations $|\alpha\>$'s on a row), we can see that $\Delta \calT$ actually does not make any contribution to the partition function. 

Although $\Delta \calT$ does not contribute to the partition function, it does have a substantial effect in the numerical simulation.
Considering from the perspective of the power method, the right dominant eigenvector of $\calT_2$ can be obtained by applying $\calT_2$ to a random vector repeatedly.
Since $\Delta \calT$ can eliminate two neighboring domain walls but lacks the ability to create any, during the power method process, the number of domain walls in the vector will keep decreasing, and the final fixed-point vector will be very close to a product state $|0\>$ with no domain walls. 
Similarly, one can infer that the left dominant eigenstate of $\calT_2$ should be close to a different product state $|\infty\>$ filled with domain walls.
Such situation is very similar to the case of the local gauge transformation discussed previously in Sec.~\ref{sec:local_gauge}, where the left and right eigenvectors are also driven towards two orthogonal product states. 

Along this line, we can further construct a gauge transformation that connects $\calT_1$ and $\calT_2$. 
We notice that $\calT_1$ can be block-diagonalized into a series of $\Uone$ blocks which correspond to different numbers of domain walls, and the additional term $\Delta \calT$ in $\calT_2$ appear as Jordan blocks between the $\Uone$ blocks. 
We consider the following gauge transformation 
\begin{equation}
    \mathcal{P} = \exp(\tau Q) = \exp(-\tau \sum_j \sigma^z_j \sigma^z_{j+1}),
    \label{eq:gauge_transformation_domain_walls}
\end{equation}
where the exponent $Q = -\sum_j \sigma_j^z \sigma_{j+1}^z$ is a diagonal Hamiltonian that measures the number of domain walls in a row of spins up to a constant.
When applying $\calP$ to the transfer matrix $\calT_2$, we obtain 
\begin{equation}
    \tilde{\calT}_2 = \calP^{-1} \calT_2 \calP = \calT_1 + \mathrm{e}^{-4\tau} \Delta \calT.
    \label{eq:gauging_T2_to_T1}
\end{equation}
After the gauge transformation, $\Delta \calT$ is suppressed exponentially with $\tau$, and when $\tau \rightarrow \infty$, the two transfer matrices become identical.
As in the case of local gauge transformations considered in Sec.~\ref{sec:local_gauge}, the gauge transformation in Eq.~\eqref{eq:gauge_transformation_domain_walls} is also an imaginary-time evolution governed by a diagonal Hamiltonian, whose ground state and highest-energy state correspond to $|0\>$ and $|\infty\>$, respectively.
From this point of view, by writing $\calT_2$ as $\calT_2 = \calP \tilde{\calT}_2 \calP^{-1}$, the failure of the boundary MPS calculation for $\calT_2$ can also be viewed as an effect of the gauge transformation. 
%When $\tau \gg 1$, $\tilde{\calT}_2$ is very close to a normal matrix, of which the left and right eigenvectors are nearly identical.
%The gauge transformation $\calP$ then drives the left and right eigenvectors towards two orthogonal product states, and suppress the physically relevant information to the tails of the entanglement spectra of the eigenvectors.

\begin{figure}[!htb]
    \centering
    \resizebox{\columnwidth}{!}{\includegraphics{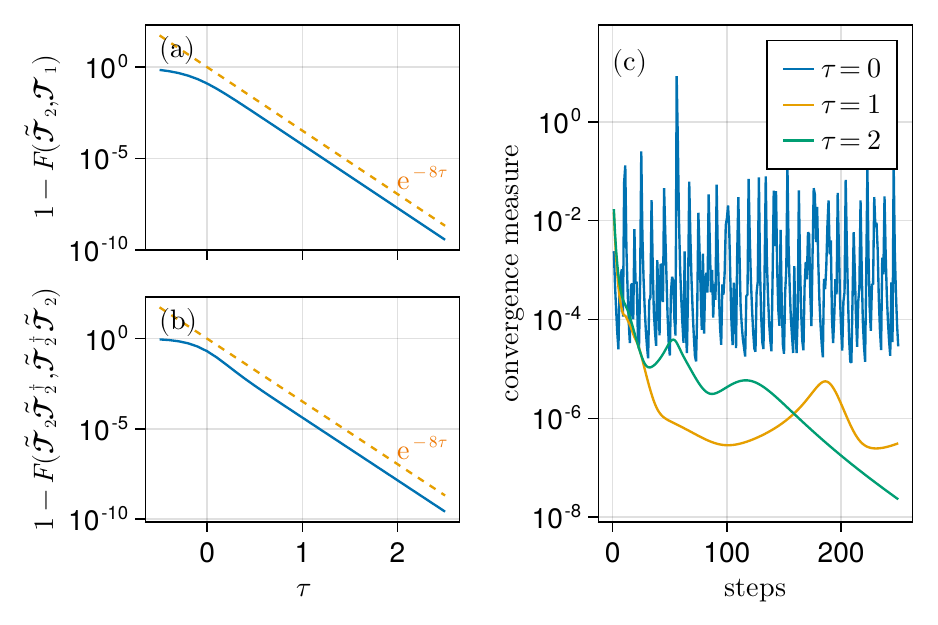}}
    \caption{Numerical results after the applying the gauge transformation to $\calT_2$.
    (a) The fidelity between $\tilde{\calT}_2$ and $\calT_1$. 
    (b) The normality of $\tilde{\calT}_2$. 
    (c) The convergence of the VUMPS before and after the gauge transformations with $\tau=1, 2$. The bond dimension of the boundary MPS is $\chi=64$.}
    \label{fig:gauging-frustrated-mpo}
\end{figure}

This gauge transformation $\calP$ can be written as a MPO with bond dimension 2, which allows us to employ this gauge transformation in the numerical calculation.
The transformed transfer matrix $\tilde{\calT}_2$ has a bond dimension of 8, but its bond dimension can be reduced to 4 without losing any precision, which is equal to the bond dimension of $\calT_1$.
We first verify Eq.~\eqref{eq:gauging_T2_to_T1} by calculating the fidelity $F(\tilde{\calT}_2, \calT_1)$ between $\tilde{\calT}_2$ and $\calT_1$.
Figure \ref{fig:gauging-frustrated-mpo} (a) shows that $1 - F(\tilde{\calT}_2, \calT_1)$ scales as $\exp(-8 \tau)$, which is consistent with Eq.~\eqref{eq:gauging_T2_to_T1}, since the overlap between $\calT_1$ and $\Delta \calT$ equals zero so that, when calculating $F(\tilde{\calT}_2, \calT_1)$, the leading correction term is proportional to $\exp(-8\tau)$. 
Accordingly, as shown in Fig.~\ref{fig:gauging-frustrated-mpo} (b), the normality measure $F(\tilde{\calT}_2\tilde{\calT}_2^\dagger, \tilde{\calT}_2^\dagger \tilde{\calT}_2)$ also goes to 1 with $\tau$ with the same scaling.
Second, we perform the VUMPS calculation with $\tilde{\calT}_2$ for different values of $\tau$. 
As shown in Fig.~\ref{fig:gauging-frustrated-mpo}~(c), the convergence of the VUMPS calculation improves with $\tau$ increasing. 
In practice, we find it not necessary to push $\tau$ to infinity, and a large enough $\tau$ is sufficient to make the VUMPS calculation converge. %, even though the $\rmU(1)$ symmetry is not completely restored in the latter case.
It is also worth noting that, in the straightforward implementation of $\tilde{\calT}_2$, the norm of the local tensor increases exponentially with $\tau$, which makes the numerical calculations unstable at large $\tau$. 
This issue can be resolved by introducing an additional gauge transformation that minimizes the norm of the local tensor in the virtual bonds of $\tilde{\calT}_2$. 

\subsection{Density matrices}
The gauge transformation $\calP$ connects the transfer matrices $\calT_1$ and $\calT_2$, which gives a natural way to interpolate between these two transfer matrices.
In the following, we calculate the density matrices $\rhoLL$, $\rhoRR$, and $\rhoLR$ for the transfer matrix $\tilde{\calT}_2$ for different choices of $\tau$.

In our calculation, the bond dimension of the boundary MPS is chosen as $\chi=64$.
For $\tau \ge 0.75$, we can directly obtain the left and right fixed-point MPSs using the power method combined with VOMPS truncation. 
For $\tau < 0.75$, we obtain the fixed-point MPSs by applying the gauge transformation $\calP$ [c.f. Eq.~\eqref{eq:gauge_transformation_domain_walls}] to the fixed-point MPSs at $\tau=0.75$, and then truncate them back to the bond dimension $\chi=32$ using VOMPS method.
Although the accuracy of power method is expected to deteriorate as $\tau$ approaches zero, we expect the results obtained below are still qualitatively correct and can be used to illustrate the effect of the gauge transformation. 

\begin{figure}[!htb]
    \centering
    \resizebox{0.667\columnwidth}{!}{\includegraphics{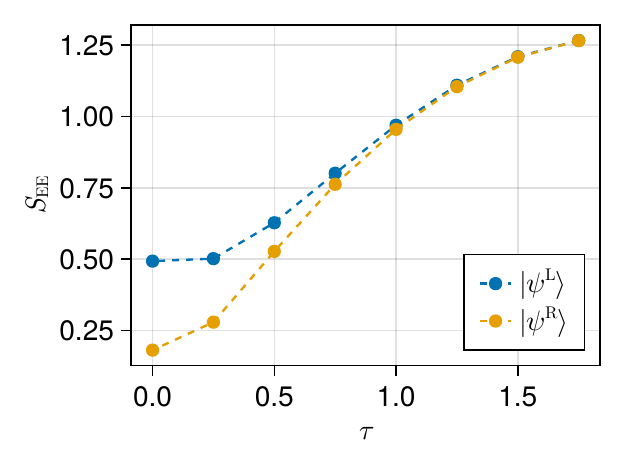}}
    \caption{The entanglement entropy of the left and right fixed-point MPSs for different choices of $\tau$.}
    \label{fig:gauging-single-side-SE}
\end{figure}

\begin{figure}[!htb]
    \centering
    \resizebox{\columnwidth}{!}{\includegraphics{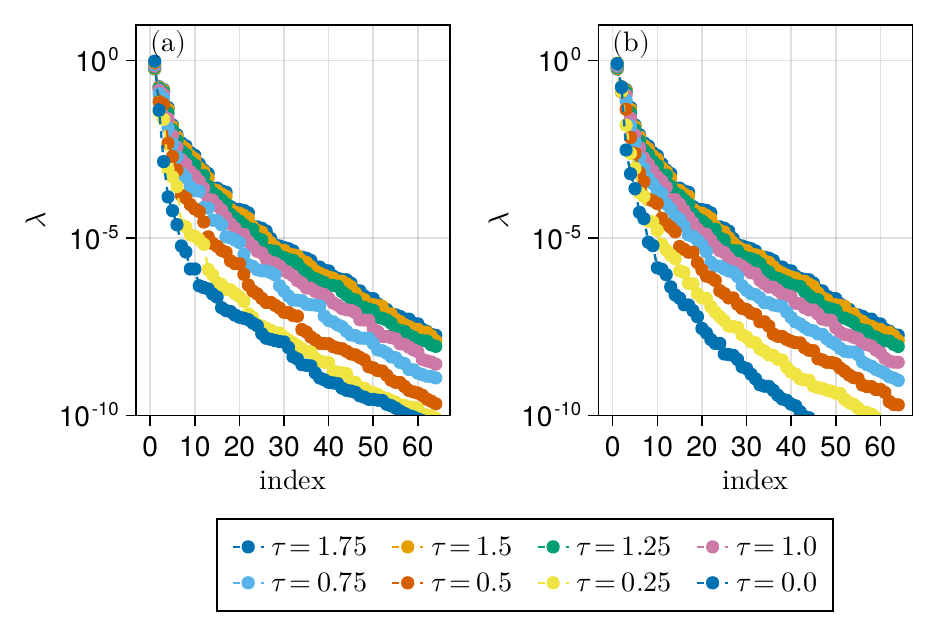}}
    \caption{The spectra of the reduced density matrix (a) $\rhoRR$ and (b) $\rhoLL$ for different choices of $\tau$.} 
    \label{fig:gauging-single-side-rho}
\end{figure}

We first show the entanglement entropy and entanglement spectrum $\rhoLL$ and $\rhoRR$ in Figs.~\ref{fig:gauging-single-side-SE} and \ref{fig:gauging-single-side-rho}. 
From the results, we see that the entanglement entropy of the left and right fixed-point MPSs decreases when the transfer matrix get closer to $\calT_2$ , i.e., when $\tau$ get closer to zero.
This is consistent with the previous analysis: the eigenstates of $\calT_2$ are very close to two orthogonal product states, which hinders the numerical calculation.  

\begin{figure}[!htb]
    \centering
    \resizebox{0.9\columnwidth}{!}{\includegraphics{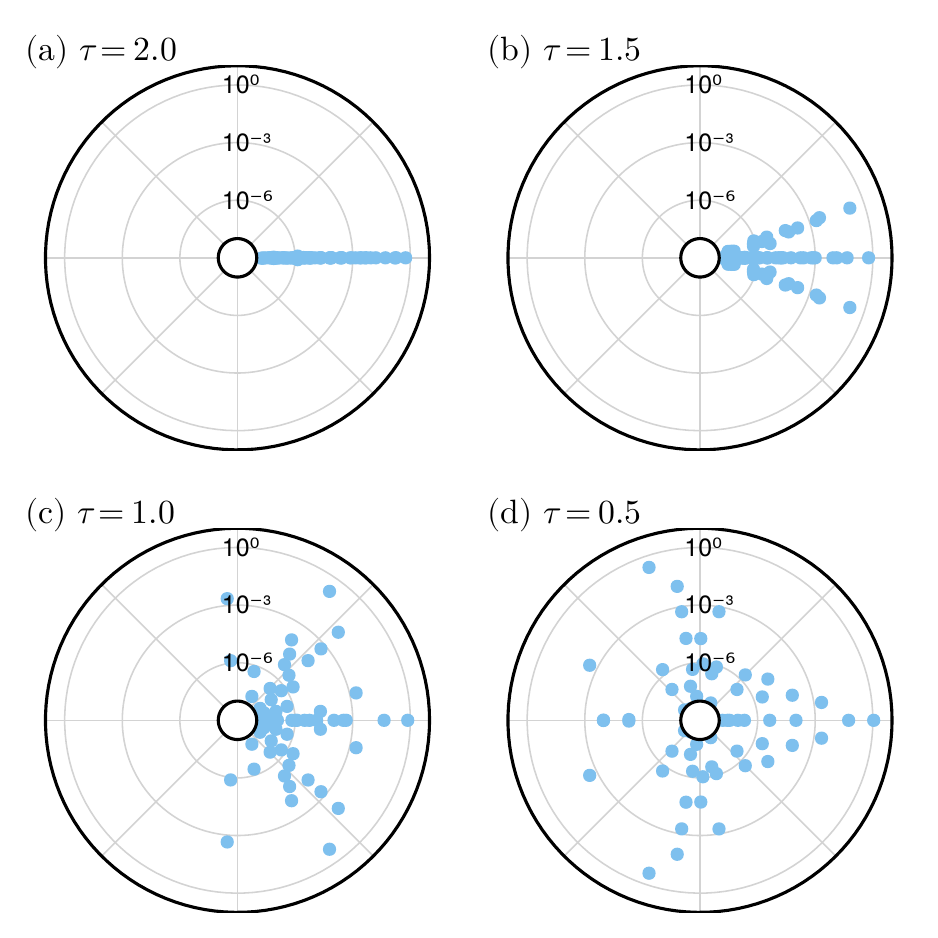}}
    \caption{Eigenspectra of the reduced density matrix $\rhoLR$ for different choices of $\tau$ plot in the complex plane. 
    The horizontal and vertical directions correspond to the real and imaginary parts of the eigenvalues, respectively.
    The radial direction in the polar plot is scaled logarithmically. 
    The eigenvalues with a norm smaller than $10^{-8}$ are not shown in the figure.} 
    \label{fig:gauging-double-side-rho}
\end{figure}

Unlike the local gauge transformations that we considered in Sec.~\ref{sec:local_gauge}, the gauge transformation \eqref{eq:gauge_transformation_domain_walls}, which is in the form of an MPO, does not preserve the spectrum of the reduced density matrix $\rhoLR$ or the (double-sided) entanglement entropy.
Using the fixed-point MPSs that we obtained, we can calculate the double-sided reduced density matrix $\rhoLR$ and its spectrum for different $\tau$'s, the result of which is shown in Fig.~\ref{fig:gauging-double-side-rho}. 

When $\tau = 2.0$, the transfer matrix is very close to the normal matrix $\calT_1$, and the eigenvalues of $\rho^\LR_A$ all locate at (or in very close vicinity of) the positive real axis.
As $\tau$ decreases, some eigenvalues deviate from the positive real axis and move towards the negative real axis. 

\begin{figure}[!htb]
    \centering
    \resizebox{\columnwidth}{!}{\includegraphics{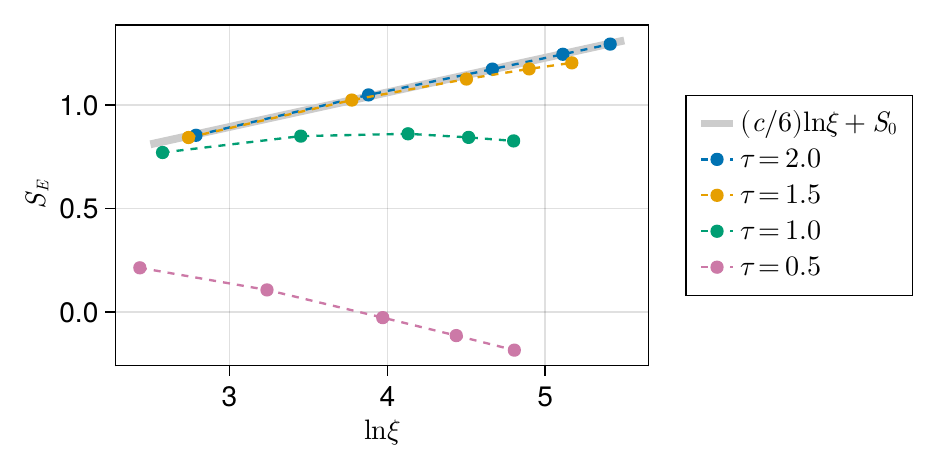}}
    \caption{Relation between the bipartite entanglement entropy $S_E$ and the correlation length $\xi$ for different choices of $\tau$.
    The light-gray solid line represents the linear relation between $S_E$ and $\ln \xi$, fitted from the data at $\tau=2.0$. 
    The central charge obtained through fitting is $c\approx 1.0028$.}
    \label{fig:gauging-double-side-SE}
\end{figure}

Since the eigenvalues of $\rho^\LR_A$ are complex numbers, in order to calculate the double-sided entanglement entropy $S^\LR_E$, we need to generalize the definition of the entanglement entropy. 
Here, we adopt the formula proposed in Ref.~\cite{tu-renyi-2022}, where the generalized entanglement entropy is defined as
\begin{equation}
    S^\LR_E = - \tr(\rho^\LR_A \ln |\rho^\LR_A|) = - \sum_i \lambda^\LR_i \ln |\lambda^\LR_i|, 
    \label{eq:double-sided-EE}
\end{equation}
where $\lambda^\LR_i$ represents the eigenvalues of $\rho^\LR_A$.
In our example the eigenvalues of $\rho^\LR_A$ always show up as conjugate pairs, so the entanglement entropy $S^\LR_E$ is real.
At different $\tau$'s, we calculate the entanglement entropy and analyze its scaling with respect to the correlation length $\xi$ for different choices of bond dimensions, the result of which is shown in Fig.~\ref{fig:gauging-single-side-SE}.
At $\tau=2.0$, where the transfer matrix is very close to the normal matrix $\calT_1$, the entanglement entropy scales linearly with the logarithm of the correlation length $\xi$.
The central charge obtained through the linear fitting is $c\approx 1.0028$, which is consistent with the prediction of the conformal field theory. 
As $\tau$ decreases, the entanglement entropy deviates from the original linear scaling relation, and the deviation becomes more significant when $\tau$ approaches zero.
When $\tau$ gets closer to zero, the entanglement entropy tend to decrease with the correlation length, and may have negative values when the correlation length is large enough.
From these results, it is clear that the entanglement properties of the degrees of freedom living on the open indices of the transfer matrix are substantially changed by the gauge transformation in the form of a nontrivial MPO.  

In the cases where the MPO deviates from a normal matrix, the current numerical results are not sufficient to determine whether there is still a linear relation between the entanglement entropy $S_E$ and $\ln \xi$ when $\xi$ is sufficiently large.
%It is also mysterious why a negative entanglement entropy can appear.
It is also unclear whether there is a physical interpretation for the negative entanglement entropy in the results.
The answer to these questions lies beyond the scope of this paper, and we leave them for future studies.

\subsection{Discussion} \label{sec:nonlocal-gauge-discussion}

In this section, we have shown that the two different tensor network representations of the partition function of the antiferromagnetic classical Ising model on the triangular lattice can be connected by a nonlocal gauge transformation represented by an MPO.
Similar to the previous example of the local gauge transformation, this nonlocal gauge transformation can also be seen as an exponential of a diagonal Hamiltonian, which alters the entanglement properties of the eigenvectors and has a substantial effect on the numerical calculation.
Unlike the local gauge transformations, the MPO gauge transformation also changes the physical entanglement properties, i.e., the entanglement between the degrees of freedom living on the open indices of the transfer matrix.

In general, if a transfer matrix can be transformed into a normal matrix by a nonlocal gauge transformation, the gauge transformation does not necessarily have the form of an exponential of a diagonal Hamiltonian, nor can it be represented by an MPO with a small bond dimension. 
Even such a MPO representation exists, it is a challenging task to find such a gauge transformation, as it is difficult to properly parametrize an MPO and its inverse.

On the other hand, we can view the difference between the two transfer matrices that we considered from a different perspective.
By examining how the ice rules are encoded into the local tensors, we can see that the spatial symmetry of the system---the $C_{6v}$ symmetry---is properly incorporated into the first tensor network construction, while only a subgroup $C_{3v}$ is locally captured in the second tensor network construction. 
The same argument applies in the example of local gauge transformations, where the local gauge transformation destroys the spatial symmetry of the original tensor network representation. 
From this point of view, both the examples that we considered demonstrate the importance of properly encoding the spatial symmetry into the local tensors when it exists, although it still remains unclear whether this is sufficient to guarantee the corresponding boundary MPS calculation is well-conditioned.

\section{Conclusion and Outlook} \label{sec:conclusion}

As a summary, in this paper, we have demonstrated that the gauge transformation on the virtual degrees of freedom of tensor networks can change the entanglement structure of the boundary MPSs and modify the way of how the physical properties are encoded into them, which can have a substantial effect on the numerical calculation.

There are still many open questions that are worth further investigation. 

First, given a transfer matrix represented by a MPO, it is not clear whether there always exists a gauge transformation that can transform it into a normal matrix.
If it exists, and it is local, as discussed in Sec.~\ref{sec:local-gauge-discussion}, one can look for it by maximizing the normality measure of the transfer matrix.
It remains an interesting open question if the same gauge transformation is obtained by imposing the minimal canonical form~\cite{acuaviva-minimal-2022}, or whether this requires additional assumptions.
More generally, if the gauge transformation is nonlocal, it is not clear how to systematically look for it, because of the difficulty in parametrizing a general MPO and its inverse.

Second, it is not clear how the discussion in this paper is related to the previous study on biorthogonalization methods.
In these methods, in the truncation process, both the left and right eigenvectors are taken into account, which largely avoids the problem which shows up in the single-directional power method that we discuss in this paper. 
In Ref.~\cite{fishman-faster-2018}, it is already demonstrated that the biorthogonalization method can successfully handle the case where the transfer matrix is connected to a normal matrix by a local gauge transformation.
However, when a nonlocal gauge transformation is involved, the double-sided reduced density matrix may be changed by the gauge transformation, and it is not clear whether the biorthogonalization method can still work in these cases. 
It also remains to be understood how the nonlocal gauge transformations affects the entanglement properties encoded in the double-sided reduced density matrix. 

Finally, as mentioned in Sec.~\ref{sec:nonlocal-gauge-discussion}, one major difference in the two tensor network representations of the partition function of the triangular-lattice antiferromagnetic Ising model is that the well-conditioned one properly incorporates the spatial symmetry of the system while the other one does not. 
This raises the question that whether properly incorporating the spatial symmetry into the tensor network representation can avoid the problem of ill-conditioning in the boundary MPS calculation.
We note that there have been studies on introducing spatial point group symmetry in practical variational calculations with PEPS, such as Refs.~\cite{jiang-symmetric-2015,mambrini-systematic-2016,hackenbroich-interplay-2018}. 
It is worth investigating which gauge transformations are still allowed after incorporating the spatial symmetries, and how will they affect the performance of the numerical calculations. 

Our code implementation is available at \href{https://github.com/tangwei94/frustrated-mpo}{\texttt{frustrated-mpo}}, which is based on the open-source package \texttt{MPSKit.jl}~\cite{Van_Damme_MPSKit_2024}.

\section*{Acknowledgment} 
We are grateful to Qi Yang, Xing-Yu Zhang, Lei Wang, Bram Vanhecke, Jeanne Colbois, Laurens Vanderstraeten, Hong-Hao Tu, Nick Bultinck, Yuan Wan, Hui-Ke Jin, Markus Scheb, Jan von Delft, Jan Schneider, Stefano Carignano, Luca Tagliacozzo, Mari Carmen Ba\~nuls, and Philippe Corboz for helpful discussions. 
This work received funding from the Research Foundation Flanders (FWO) via grant GOE1520N.

%\appendix
%\section{appendix1}\label{app:temp1}

\bibliography{nonHermitian}

\begin{thebibliography}{61}%
\makeatletter
\providecommand \@ifxundefined [1]{%
 \@ifx{#1\undefined}
}%
\providecommand \@ifnum [1]{%
 \ifnum #1\expandafter \@firstoftwo
 \else \expandafter \@secondoftwo
 \fi
}%
\providecommand \@ifx [1]{%
 \ifx #1\expandafter \@firstoftwo
 \else \expandafter \@secondoftwo
 \fi
}%
\providecommand \natexlab [1]{#1}%
\providecommand \enquote  [1]{``#1''}%
\providecommand \bibnamefont  [1]{#1}%
\providecommand \bibfnamefont [1]{#1}%
\providecommand \citenamefont [1]{#1}%
\providecommand \href@noop [0]{\@secondoftwo}%
\providecommand \href [0]{\begingroup \@sanitize@url \@href}%
\providecommand \@href[1]{\@@startlink{#1}\@@href}%
\providecommand \@@href[1]{\endgroup#1\@@endlink}%
\providecommand \@sanitize@url [0]{\catcode `\\12\catcode `\$12\catcode `\&12\catcode `\#12\catcode `\^12\catcode `\_12\catcode `\%12\relax}%
\providecommand \@@startlink[1]{}%
\providecommand \@@endlink[0]{}%
\providecommand \url  [0]{\begingroup\@sanitize@url \@url }%
\providecommand \@url [1]{\endgroup\@href {#1}{\urlprefix }}%
\providecommand \urlprefix  [0]{URL }%
\providecommand \Eprint [0]{\href }%
\providecommand \doibase [0]{https://doi.org/}%
\providecommand \selectlanguage [0]{\@gobble}%
\providecommand \bibinfo  [0]{\@secondoftwo}%
\providecommand \bibfield  [0]{\@secondoftwo}%
\providecommand \translation [1]{[#1]}%
\providecommand \BibitemOpen [0]{}%
\providecommand \bibitemStop [0]{}%
\providecommand \bibitemNoStop [0]{.\EOS\space}%
\providecommand \EOS [0]{\spacefactor3000\relax}%
\providecommand \BibitemShut  [1]{\csname bibitem#1\endcsname}%
\let\auto@bib@innerbib\@empty
%</preamble>
\bibitem [{\citenamefont {Schollw{\"o}ck}(2011)}]{schollwock-density-2011}%
  \BibitemOpen
  \bibfield  {author} {\bibinfo {author} {\bibfnamefont {U.}~\bibnamefont {Schollw{\"o}ck}},\ }\bibfield  {title} {\bibinfo {title} {The density-matrix renormalization group in the age of matrix product states},\ }\href {https://www.sciencedirect.com/science/article/pii/S0003491610001752} {\bibfield  {journal} {\bibinfo  {journal} {Ann. Phys.}\ }\textbf {\bibinfo {volume} {326}},\ \bibinfo {pages} {96} (\bibinfo {year} {2011})}\BibitemShut {NoStop}%
\bibitem [{\citenamefont {Or\'us}(2014)}]{orus-practical-2014}%
  \BibitemOpen
  \bibfield  {author} {\bibinfo {author} {\bibfnamefont {R.}~\bibnamefont {Or\'us}},\ }\bibfield  {title} {\bibinfo {title} {A practical introduction to tensor networks: Matrix product states and projected entangled pair states},\ }\href {https://doi.org/https://doi.org/10.1016/j.aop.2014.06.013} {\bibfield  {journal} {\bibinfo  {journal} {Ann. Phys.}\ }\textbf {\bibinfo {volume} {349}},\ \bibinfo {pages} {117} (\bibinfo {year} {2014})}\BibitemShut {NoStop}%
\bibitem [{\citenamefont {Cirac}\ \emph {et~al.}(2021)\citenamefont {Cirac}, \citenamefont {P\'erez-Garc\'{\i}a}, \citenamefont {Schuch},\ and\ \citenamefont {Verstraete}}]{cirac-matrix-2021}%
  \BibitemOpen
  \bibfield  {author} {\bibinfo {author} {\bibfnamefont {J.~I.}\ \bibnamefont {Cirac}}, \bibinfo {author} {\bibfnamefont {D.}~\bibnamefont {P\'erez-Garc\'{\i}a}}, \bibinfo {author} {\bibfnamefont {N.}~\bibnamefont {Schuch}},\ and\ \bibinfo {author} {\bibfnamefont {F.}~\bibnamefont {Verstraete}},\ }\bibfield  {title} {\bibinfo {title} {Matrix product states and projected entangled pair states: Concepts, symmetries, theorems},\ }\href {https://doi.org/10.1103/RevModPhys.93.045003} {\bibfield  {journal} {\bibinfo  {journal} {Rev. Mod. Phys.}\ }\textbf {\bibinfo {volume} {93}},\ \bibinfo {pages} {045003} (\bibinfo {year} {2021})}\BibitemShut {NoStop}%
\bibitem [{\citenamefont {Xiang}(2023)}]{xiang-density-book-2023}%
  \BibitemOpen
  \bibfield  {author} {\bibinfo {author} {\bibfnamefont {T.}~\bibnamefont {Xiang}},\ }\href {https://www.cambridge.org/be/universitypress/subjects/physics/condensed-matter-physics-nanoscience-and-mesoscopic-physics/density-matrix-and-tensor-network-renormalization?format=HB} {\emph {\bibinfo {title} {Density Matrix and Tensor Network Renormalization}}}\ (\bibinfo  {publisher} {Cambridge University Press},\ \bibinfo {year} {2023})\BibitemShut {NoStop}%
\bibitem [{\citenamefont {Nishino}\ and\ \citenamefont {Okunishi}(1996)}]{nishino-corner-1996}%
  \BibitemOpen
  \bibfield  {author} {\bibinfo {author} {\bibfnamefont {T.}~\bibnamefont {Nishino}}\ and\ \bibinfo {author} {\bibfnamefont {K.}~\bibnamefont {Okunishi}},\ }\bibfield  {title} {\bibinfo {title} {Corner transfer matrix renormalization group method},\ }\href {https://journals.jps.jp/doi/pdf/10.1143/JPSJ.65.891} {\bibfield  {journal} {\bibinfo  {journal} {J. Phys. Soc. Jpn}\ }\textbf {\bibinfo {volume} {65}},\ \bibinfo {pages} {891} (\bibinfo {year} {1996})}\BibitemShut {NoStop}%
\bibitem [{\citenamefont {Nishino}\ and\ \citenamefont {Okunishi}(1997)}]{nishino-corner-1997}%
  \BibitemOpen
  \bibfield  {author} {\bibinfo {author} {\bibfnamefont {T.}~\bibnamefont {Nishino}}\ and\ \bibinfo {author} {\bibfnamefont {K.}~\bibnamefont {Okunishi}},\ }\bibfield  {title} {\bibinfo {title} {Corner transfer matrix algorithm for classical renormalization group},\ }\href {https://journals.jps.jp/doi/10.1143/JPSJ.66.3040} {\bibfield  {journal} {\bibinfo  {journal} {J. Phys. Soc. Jpn}\ }\textbf {\bibinfo {volume} {66}},\ \bibinfo {pages} {3040} (\bibinfo {year} {1997})}\BibitemShut {NoStop}%
\bibitem [{\citenamefont {Or\'us}\ and\ \citenamefont {Vidal}(2009)}]{orus-simulation-2009}%
  \BibitemOpen
  \bibfield  {author} {\bibinfo {author} {\bibfnamefont {R.}~\bibnamefont {Or\'us}}\ and\ \bibinfo {author} {\bibfnamefont {G.}~\bibnamefont {Vidal}},\ }\bibfield  {title} {\bibinfo {title} {Simulation of two-dimensional quantum systems on an infinite lattice revisited: Corner transfer matrix for tensor contraction},\ }\href {https://doi.org/10.1103/PhysRevB.80.094403} {\bibfield  {journal} {\bibinfo  {journal} {Phys. Rev. B}\ }\textbf {\bibinfo {volume} {80}},\ \bibinfo {pages} {094403} (\bibinfo {year} {2009})}\BibitemShut {NoStop}%
\bibitem [{\citenamefont {Corboz}\ \emph {et~al.}(2010)\citenamefont {Corboz}, \citenamefont {Jordan},\ and\ \citenamefont {Vidal}}]{corboz-simulation-2010}%
  \BibitemOpen
  \bibfield  {author} {\bibinfo {author} {\bibfnamefont {P.}~\bibnamefont {Corboz}}, \bibinfo {author} {\bibfnamefont {J.}~\bibnamefont {Jordan}},\ and\ \bibinfo {author} {\bibfnamefont {G.}~\bibnamefont {Vidal}},\ }\bibfield  {title} {\bibinfo {title} {Simulation of fermionic lattice models in two dimensions with projected entangled-pair states: Next-nearest neighbor hamiltonians},\ }\href {https://doi.org/10.1103/PhysRevB.82.245119} {\bibfield  {journal} {\bibinfo  {journal} {Phys. Rev. B}\ }\textbf {\bibinfo {volume} {82}},\ \bibinfo {pages} {245119} (\bibinfo {year} {2010})}\BibitemShut {NoStop}%
\bibitem [{\citenamefont {Or\'us}(2012)}]{orus-exploring-2012}%
  \BibitemOpen
  \bibfield  {author} {\bibinfo {author} {\bibfnamefont {R.}~\bibnamefont {Or\'us}},\ }\bibfield  {title} {\bibinfo {title} {Exploring corner transfer matrices and corner tensors for the classical simulation of quantum lattice systems},\ }\href {https://doi.org/10.1103/PhysRevB.85.205117} {\bibfield  {journal} {\bibinfo  {journal} {Phys. Rev. B}\ }\textbf {\bibinfo {volume} {85}},\ \bibinfo {pages} {205117} (\bibinfo {year} {2012})}\BibitemShut {NoStop}%
\bibitem [{\citenamefont {Corboz}\ \emph {et~al.}(2014)\citenamefont {Corboz}, \citenamefont {Rice},\ and\ \citenamefont {Troyer}}]{corboz-competing-2014}%
  \BibitemOpen
  \bibfield  {author} {\bibinfo {author} {\bibfnamefont {P.}~\bibnamefont {Corboz}}, \bibinfo {author} {\bibfnamefont {T.~M.}\ \bibnamefont {Rice}},\ and\ \bibinfo {author} {\bibfnamefont {M.}~\bibnamefont {Troyer}},\ }\bibfield  {title} {\bibinfo {title} {Competing states in the $t$-$j$ model: Uniform $d$-wave state versus stripe state},\ }\href {https://doi.org/10.1103/PhysRevLett.113.046402} {\bibfield  {journal} {\bibinfo  {journal} {Phys. Rev. Lett.}\ }\textbf {\bibinfo {volume} {113}},\ \bibinfo {pages} {046402} (\bibinfo {year} {2014})}\BibitemShut {NoStop}%
\bibitem [{\citenamefont {Levin}\ and\ \citenamefont {Nave}(2007)}]{levin-tensor-2007}%
  \BibitemOpen
  \bibfield  {author} {\bibinfo {author} {\bibfnamefont {M.}~\bibnamefont {Levin}}\ and\ \bibinfo {author} {\bibfnamefont {C.~P.}\ \bibnamefont {Nave}},\ }\bibfield  {title} {\bibinfo {title} {Tensor renormalization group approach to two-dimensional classical lattice models},\ }\href {https://doi.org/10.1103/PhysRevLett.99.120601} {\bibfield  {journal} {\bibinfo  {journal} {Phys. Rev. Lett.}\ }\textbf {\bibinfo {volume} {99}},\ \bibinfo {pages} {120601} (\bibinfo {year} {2007})}\BibitemShut {NoStop}%
\bibitem [{\citenamefont {Xie}\ \emph {et~al.}(2009)\citenamefont {Xie}, \citenamefont {Jiang}, \citenamefont {Chen}, \citenamefont {Weng},\ and\ \citenamefont {Xiang}}]{xie-second-2009}%
  \BibitemOpen
  \bibfield  {author} {\bibinfo {author} {\bibfnamefont {Z.~Y.}\ \bibnamefont {Xie}}, \bibinfo {author} {\bibfnamefont {H.~C.}\ \bibnamefont {Jiang}}, \bibinfo {author} {\bibfnamefont {Q.~N.}\ \bibnamefont {Chen}}, \bibinfo {author} {\bibfnamefont {Z.~Y.}\ \bibnamefont {Weng}},\ and\ \bibinfo {author} {\bibfnamefont {T.}~\bibnamefont {Xiang}},\ }\bibfield  {title} {\bibinfo {title} {Second renormalization of tensor-network states},\ }\href {https://doi.org/10.1103/PhysRevLett.103.160601} {\bibfield  {journal} {\bibinfo  {journal} {Phys. Rev. Lett.}\ }\textbf {\bibinfo {volume} {103}},\ \bibinfo {pages} {160601} (\bibinfo {year} {2009})}\BibitemShut {NoStop}%
\bibitem [{\citenamefont {Gu}\ and\ \citenamefont {Wen}(2009)}]{gu-tensor-2009}%
  \BibitemOpen
  \bibfield  {author} {\bibinfo {author} {\bibfnamefont {Z.-C.}\ \bibnamefont {Gu}}\ and\ \bibinfo {author} {\bibfnamefont {X.-G.}\ \bibnamefont {Wen}},\ }\bibfield  {title} {\bibinfo {title} {Tensor-entanglement-filtering renormalization approach and symmetry-protected topological order},\ }\href {https://doi.org/10.1103/PhysRevB.80.155131} {\bibfield  {journal} {\bibinfo  {journal} {Phys. Rev. B}\ }\textbf {\bibinfo {volume} {80}},\ \bibinfo {pages} {155131} (\bibinfo {year} {2009})}\BibitemShut {NoStop}%
\bibitem [{\citenamefont {Li}\ \emph {et~al.}(2011)\citenamefont {Li}, \citenamefont {Ran}, \citenamefont {Gong}, \citenamefont {Zhao}, \citenamefont {Xi}, \citenamefont {Ye},\ and\ \citenamefont {Su}}]{li-linearized-2011}%
  \BibitemOpen
  \bibfield  {author} {\bibinfo {author} {\bibfnamefont {W.}~\bibnamefont {Li}}, \bibinfo {author} {\bibfnamefont {S.-J.}\ \bibnamefont {Ran}}, \bibinfo {author} {\bibfnamefont {S.-S.}\ \bibnamefont {Gong}}, \bibinfo {author} {\bibfnamefont {Y.}~\bibnamefont {Zhao}}, \bibinfo {author} {\bibfnamefont {B.}~\bibnamefont {Xi}}, \bibinfo {author} {\bibfnamefont {F.}~\bibnamefont {Ye}},\ and\ \bibinfo {author} {\bibfnamefont {G.}~\bibnamefont {Su}},\ }\bibfield  {title} {\bibinfo {title} {Linearized tensor renormalization group algorithm for the calculation of thermodynamic properties of quantum lattice models},\ }\href {https://doi.org/10.1103/PhysRevLett.106.127202} {\bibfield  {journal} {\bibinfo  {journal} {Phys. Rev. Lett.}\ }\textbf {\bibinfo {volume} {106}},\ \bibinfo {pages} {127202} (\bibinfo {year} {2011})}\BibitemShut {NoStop}%
\bibitem [{\citenamefont {Xie}\ \emph {et~al.}(2012)\citenamefont {Xie}, \citenamefont {Chen}, \citenamefont {Qin}, \citenamefont {Zhu}, \citenamefont {Yang},\ and\ \citenamefont {Xiang}}]{xie-coarse-2012}%
  \BibitemOpen
  \bibfield  {author} {\bibinfo {author} {\bibfnamefont {Z.~Y.}\ \bibnamefont {Xie}}, \bibinfo {author} {\bibfnamefont {J.}~\bibnamefont {Chen}}, \bibinfo {author} {\bibfnamefont {M.~P.}\ \bibnamefont {Qin}}, \bibinfo {author} {\bibfnamefont {J.~W.}\ \bibnamefont {Zhu}}, \bibinfo {author} {\bibfnamefont {L.~P.}\ \bibnamefont {Yang}},\ and\ \bibinfo {author} {\bibfnamefont {T.}~\bibnamefont {Xiang}},\ }\bibfield  {title} {\bibinfo {title} {Coarse-graining renormalization by higher-order singular value decomposition},\ }\href {https://doi.org/10.1103/PhysRevB.86.045139} {\bibfield  {journal} {\bibinfo  {journal} {Phys. Rev. B}\ }\textbf {\bibinfo {volume} {86}},\ \bibinfo {pages} {045139} (\bibinfo {year} {2012})}\BibitemShut {NoStop}%
\bibitem [{\citenamefont {Evenbly}\ and\ \citenamefont {Vidal}(2015)}]{evenbly-tensor-2015}%
  \BibitemOpen
  \bibfield  {author} {\bibinfo {author} {\bibfnamefont {G.}~\bibnamefont {Evenbly}}\ and\ \bibinfo {author} {\bibfnamefont {G.}~\bibnamefont {Vidal}},\ }\bibfield  {title} {\bibinfo {title} {Tensor network renormalization},\ }\href {https://doi.org/10.1103/PhysRevLett.115.180405} {\bibfield  {journal} {\bibinfo  {journal} {Phys. Rev. Lett.}\ }\textbf {\bibinfo {volume} {115}},\ \bibinfo {pages} {180405} (\bibinfo {year} {2015})}\BibitemShut {NoStop}%
\bibitem [{\citenamefont {Czarnik}\ and\ \citenamefont {Dziarmaga}(2015)}]{czarnik-variational-2015}%
  \BibitemOpen
  \bibfield  {author} {\bibinfo {author} {\bibfnamefont {P.}~\bibnamefont {Czarnik}}\ and\ \bibinfo {author} {\bibfnamefont {J.}~\bibnamefont {Dziarmaga}},\ }\bibfield  {title} {\bibinfo {title} {Variational approach to projected entangled pair states at finite temperature},\ }\href {https://doi.org/10.1103/PhysRevB.92.035152} {\bibfield  {journal} {\bibinfo  {journal} {Phys. Rev. B}\ }\textbf {\bibinfo {volume} {92}},\ \bibinfo {pages} {035152} (\bibinfo {year} {2015})}\BibitemShut {NoStop}%
\bibitem [{\citenamefont {Yang}\ \emph {et~al.}(2017)\citenamefont {Yang}, \citenamefont {Gu},\ and\ \citenamefont {Wen}}]{yang-loop-2017}%
  \BibitemOpen
  \bibfield  {author} {\bibinfo {author} {\bibfnamefont {S.}~\bibnamefont {Yang}}, \bibinfo {author} {\bibfnamefont {Z.-C.}\ \bibnamefont {Gu}},\ and\ \bibinfo {author} {\bibfnamefont {X.-G.}\ \bibnamefont {Wen}},\ }\bibfield  {title} {\bibinfo {title} {Loop optimization for tensor network renormalization},\ }\href {https://doi.org/10.1103/PhysRevLett.118.110504} {\bibfield  {journal} {\bibinfo  {journal} {Phys. Rev. Lett.}\ }\textbf {\bibinfo {volume} {118}},\ \bibinfo {pages} {110504} (\bibinfo {year} {2017})}\BibitemShut {NoStop}%
\bibitem [{\citenamefont {Bal}\ \emph {et~al.}(2017)\citenamefont {Bal}, \citenamefont {Mari\"en}, \citenamefont {Haegeman},\ and\ \citenamefont {Verstraete}}]{bal-renormalization-2017}%
  \BibitemOpen
  \bibfield  {author} {\bibinfo {author} {\bibfnamefont {M.}~\bibnamefont {Bal}}, \bibinfo {author} {\bibfnamefont {M.}~\bibnamefont {Mari\"en}}, \bibinfo {author} {\bibfnamefont {J.}~\bibnamefont {Haegeman}},\ and\ \bibinfo {author} {\bibfnamefont {F.}~\bibnamefont {Verstraete}},\ }\bibfield  {title} {\bibinfo {title} {Renormalization group flows of hamiltonians using tensor networks},\ }\href {https://doi.org/10.1103/PhysRevLett.118.250602} {\bibfield  {journal} {\bibinfo  {journal} {Phys. Rev. Lett.}\ }\textbf {\bibinfo {volume} {118}},\ \bibinfo {pages} {250602} (\bibinfo {year} {2017})}\BibitemShut {NoStop}%
\bibitem [{\citenamefont {Hauru}\ \emph {et~al.}(2018)\citenamefont {Hauru}, \citenamefont {Delcamp},\ and\ \citenamefont {Mizera}}]{hauru-renormalization-2018}%
  \BibitemOpen
  \bibfield  {author} {\bibinfo {author} {\bibfnamefont {M.}~\bibnamefont {Hauru}}, \bibinfo {author} {\bibfnamefont {C.}~\bibnamefont {Delcamp}},\ and\ \bibinfo {author} {\bibfnamefont {S.}~\bibnamefont {Mizera}},\ }\bibfield  {title} {\bibinfo {title} {Renormalization of tensor networks using graph-independent local truncations},\ }\href {https://doi.org/10.1103/PhysRevB.97.045111} {\bibfield  {journal} {\bibinfo  {journal} {Phys. Rev. B}\ }\textbf {\bibinfo {volume} {97}},\ \bibinfo {pages} {045111} (\bibinfo {year} {2018})}\BibitemShut {NoStop}%
\bibitem [{\citenamefont {Chen}\ \emph {et~al.}(2018)\citenamefont {Chen}, \citenamefont {Chen}, \citenamefont {Chen}, \citenamefont {Li},\ and\ \citenamefont {Weichselbaum}}]{chen-exponential-2018}%
  \BibitemOpen
  \bibfield  {author} {\bibinfo {author} {\bibfnamefont {B.-B.}\ \bibnamefont {Chen}}, \bibinfo {author} {\bibfnamefont {L.}~\bibnamefont {Chen}}, \bibinfo {author} {\bibfnamefont {Z.}~\bibnamefont {Chen}}, \bibinfo {author} {\bibfnamefont {W.}~\bibnamefont {Li}},\ and\ \bibinfo {author} {\bibfnamefont {A.}~\bibnamefont {Weichselbaum}},\ }\bibfield  {title} {\bibinfo {title} {Exponential thermal tensor network approach for quantum lattice models},\ }\href {https://doi.org/10.1103/PhysRevX.8.031082} {\bibfield  {journal} {\bibinfo  {journal} {Phys. Rev. X}\ }\textbf {\bibinfo {volume} {8}},\ \bibinfo {pages} {031082} (\bibinfo {year} {2018})}\BibitemShut {NoStop}%
\bibitem [{\citenamefont {Nishino}(1995)}]{nishino-density-1995}%
  \BibitemOpen
  \bibfield  {author} {\bibinfo {author} {\bibfnamefont {T.}~\bibnamefont {Nishino}},\ }\bibfield  {title} {\bibinfo {title} {Density matrix renormalization group method for 2d classical models},\ }\href {https://journals.jps.jp/doi/10.1143/JPSJ.64.3598} {\bibfield  {journal} {\bibinfo  {journal} {J. Phys. Soc. Jpn}\ }\textbf {\bibinfo {volume} {64}},\ \bibinfo {pages} {3598} (\bibinfo {year} {1995})}\BibitemShut {NoStop}%
\bibitem [{\citenamefont {Wang}\ and\ \citenamefont {Xiang}(1997)}]{wang-transfer-1997}%
  \BibitemOpen
  \bibfield  {author} {\bibinfo {author} {\bibfnamefont {X.}~\bibnamefont {Wang}}\ and\ \bibinfo {author} {\bibfnamefont {T.}~\bibnamefont {Xiang}},\ }\bibfield  {title} {\bibinfo {title} {Transfer-matrix density-matrix renormalization-group theory for thermodynamics of one-dimensional quantum systems},\ }\href {https://doi.org/10.1103/PhysRevB.56.5061} {\bibfield  {journal} {\bibinfo  {journal} {Phys. Rev. B}\ }\textbf {\bibinfo {volume} {56}},\ \bibinfo {pages} {5061} (\bibinfo {year} {1997})}\BibitemShut {NoStop}%
\bibitem [{\citenamefont {Xiang}(1998)}]{xiang-thermodynamics-1998}%
  \BibitemOpen
  \bibfield  {author} {\bibinfo {author} {\bibfnamefont {T.}~\bibnamefont {Xiang}},\ }\bibfield  {title} {\bibinfo {title} {Thermodynamics of quantum heisenberg spin chains},\ }\href {https://doi.org/10.1103/PhysRevB.58.9142} {\bibfield  {journal} {\bibinfo  {journal} {Phys. Rev. B}\ }\textbf {\bibinfo {volume} {58}},\ \bibinfo {pages} {9142} (\bibinfo {year} {1998})}\BibitemShut {NoStop}%
\bibitem [{\citenamefont {Vidal}(2003)}]{vidal-efficient-2003}%
  \BibitemOpen
  \bibfield  {author} {\bibinfo {author} {\bibfnamefont {G.}~\bibnamefont {Vidal}},\ }\bibfield  {title} {\bibinfo {title} {Efficient classical simulation of slightly entangled quantum computations},\ }\href {https://doi.org/10.1103/PhysRevLett.91.147902} {\bibfield  {journal} {\bibinfo  {journal} {Phys. Rev. Lett.}\ }\textbf {\bibinfo {volume} {91}},\ \bibinfo {pages} {147902} (\bibinfo {year} {2003})}\BibitemShut {NoStop}%
\bibitem [{\citenamefont {Or\'us}\ and\ \citenamefont {Vidal}(2008)}]{orus-infinite-2008}%
  \BibitemOpen
  \bibfield  {author} {\bibinfo {author} {\bibfnamefont {R.}~\bibnamefont {Or\'us}}\ and\ \bibinfo {author} {\bibfnamefont {G.}~\bibnamefont {Vidal}},\ }\bibfield  {title} {\bibinfo {title} {Infinite time-evolving block decimation algorithm beyond unitary evolution},\ }\href {https://doi.org/10.1103/PhysRevB.78.155117} {\bibfield  {journal} {\bibinfo  {journal} {Phys. Rev. B}\ }\textbf {\bibinfo {volume} {78}},\ \bibinfo {pages} {155117} (\bibinfo {year} {2008})}\BibitemShut {NoStop}%
\bibitem [{\citenamefont {Jordan}\ \emph {et~al.}(2008)\citenamefont {Jordan}, \citenamefont {Or\'us}, \citenamefont {Vidal}, \citenamefont {Verstraete},\ and\ \citenamefont {Cirac}}]{jordan-classical-2008}%
  \BibitemOpen
  \bibfield  {author} {\bibinfo {author} {\bibfnamefont {J.}~\bibnamefont {Jordan}}, \bibinfo {author} {\bibfnamefont {R.}~\bibnamefont {Or\'us}}, \bibinfo {author} {\bibfnamefont {G.}~\bibnamefont {Vidal}}, \bibinfo {author} {\bibfnamefont {F.}~\bibnamefont {Verstraete}},\ and\ \bibinfo {author} {\bibfnamefont {J.~I.}\ \bibnamefont {Cirac}},\ }\bibfield  {title} {\bibinfo {title} {Classical simulation of infinite-size quantum lattice systems in two spatial dimensions},\ }\href {https://doi.org/10.1103/PhysRevLett.101.250602} {\bibfield  {journal} {\bibinfo  {journal} {Phys. Rev. Lett.}\ }\textbf {\bibinfo {volume} {101}},\ \bibinfo {pages} {250602} (\bibinfo {year} {2008})}\BibitemShut {NoStop}%
\bibitem [{\citenamefont {Zauner-Stauber}\ \emph {et~al.}(2018)\citenamefont {Zauner-Stauber}, \citenamefont {Vanderstraeten}, \citenamefont {Fishman}, \citenamefont {Verstraete},\ and\ \citenamefont {Haegeman}}]{zauner-stauber-variational-2018}%
  \BibitemOpen
  \bibfield  {author} {\bibinfo {author} {\bibfnamefont {V.}~\bibnamefont {Zauner-Stauber}}, \bibinfo {author} {\bibfnamefont {L.}~\bibnamefont {Vanderstraeten}}, \bibinfo {author} {\bibfnamefont {M.~T.}\ \bibnamefont {Fishman}}, \bibinfo {author} {\bibfnamefont {F.}~\bibnamefont {Verstraete}},\ and\ \bibinfo {author} {\bibfnamefont {J.}~\bibnamefont {Haegeman}},\ }\bibfield  {title} {\bibinfo {title} {Variational optimization algorithms for uniform matrix product states},\ }\href {https://doi.org/10.1103/PhysRevB.97.045145} {\bibfield  {journal} {\bibinfo  {journal} {Phys. Rev. B}\ }\textbf {\bibinfo {volume} {97}},\ \bibinfo {pages} {045145} (\bibinfo {year} {2018})}\BibitemShut {NoStop}%
\bibitem [{\citenamefont {Haegeman}\ \emph {et~al.}(2016)\citenamefont {Haegeman}, \citenamefont {Lubich}, \citenamefont {Oseledets}, \citenamefont {Vandereycken},\ and\ \citenamefont {Verstraete}}]{haegeman-unifying-2016}%
  \BibitemOpen
  \bibfield  {author} {\bibinfo {author} {\bibfnamefont {J.}~\bibnamefont {Haegeman}}, \bibinfo {author} {\bibfnamefont {C.}~\bibnamefont {Lubich}}, \bibinfo {author} {\bibfnamefont {I.}~\bibnamefont {Oseledets}}, \bibinfo {author} {\bibfnamefont {B.}~\bibnamefont {Vandereycken}},\ and\ \bibinfo {author} {\bibfnamefont {F.}~\bibnamefont {Verstraete}},\ }\bibfield  {title} {\bibinfo {title} {Unifying time evolution and optimization with matrix product states},\ }\href {https://doi.org/10.1103/PhysRevB.94.165116} {\bibfield  {journal} {\bibinfo  {journal} {Phys. Rev. B}\ }\textbf {\bibinfo {volume} {94}},\ \bibinfo {pages} {165116} (\bibinfo {year} {2016})}\BibitemShut {NoStop}%
\bibitem [{\citenamefont {Haegeman}\ and\ \citenamefont {Verstraete}(2017)}]{haegeman-diagonalizing-2017}%
  \BibitemOpen
  \bibfield  {author} {\bibinfo {author} {\bibfnamefont {J.}~\bibnamefont {Haegeman}}\ and\ \bibinfo {author} {\bibfnamefont {F.}~\bibnamefont {Verstraete}},\ }\bibfield  {title} {\bibinfo {title} {Diagonalizing transfer matrices and matrix product operators: A medley of exact and computational methods},\ }\href {https://doi.org/10.1146/annurev-conmatphys-031016-025507} {\bibfield  {journal} {\bibinfo  {journal} {Annu. Rev. Condens. Matter Phys.}\ }\textbf {\bibinfo {volume} {8}},\ \bibinfo {pages} {355} (\bibinfo {year} {2017})}\BibitemShut {NoStop}%
\bibitem [{\citenamefont {Vanderstraeten}\ \emph {et~al.}(2019)\citenamefont {Vanderstraeten}, \citenamefont {Haegeman},\ and\ \citenamefont {Verstraete}}]{vanderstraeten-tangent-2019}%
  \BibitemOpen
  \bibfield  {author} {\bibinfo {author} {\bibfnamefont {L.}~\bibnamefont {Vanderstraeten}}, \bibinfo {author} {\bibfnamefont {J.}~\bibnamefont {Haegeman}},\ and\ \bibinfo {author} {\bibfnamefont {F.}~\bibnamefont {Verstraete}},\ }\bibfield  {title} {\bibinfo {title} {Tangent-space methods for uniform matrix product states},\ }\href {https://scipost.org/10.21468/SciPostPhysLectNotes.7} {\bibfield  {journal} {\bibinfo  {journal} {SciPost Phys. Lect. Notes}\ } (\bibinfo {year} {2019})}\BibitemShut {NoStop}%
\bibitem [{\citenamefont {Vanderstraeten}\ \emph {et~al.}(2022)\citenamefont {Vanderstraeten}, \citenamefont {Burgelman}, \citenamefont {Ponsioen}, \citenamefont {Van~Damme}, \citenamefont {Vanhecke}, \citenamefont {Corboz}, \citenamefont {Haegeman},\ and\ \citenamefont {Verstraete}}]{vanderstraeten-variational-2022}%
  \BibitemOpen
  \bibfield  {author} {\bibinfo {author} {\bibfnamefont {L.}~\bibnamefont {Vanderstraeten}}, \bibinfo {author} {\bibfnamefont {L.}~\bibnamefont {Burgelman}}, \bibinfo {author} {\bibfnamefont {B.}~\bibnamefont {Ponsioen}}, \bibinfo {author} {\bibfnamefont {M.}~\bibnamefont {Van~Damme}}, \bibinfo {author} {\bibfnamefont {B.}~\bibnamefont {Vanhecke}}, \bibinfo {author} {\bibfnamefont {P.}~\bibnamefont {Corboz}}, \bibinfo {author} {\bibfnamefont {J.}~\bibnamefont {Haegeman}},\ and\ \bibinfo {author} {\bibfnamefont {F.}~\bibnamefont {Verstraete}},\ }\bibfield  {title} {\bibinfo {title} {Variational methods for contracting projected entangled-pair states},\ }\href {https://doi.org/10.1103/PhysRevB.105.195140} {\bibfield  {journal} {\bibinfo  {journal} {Phys. Rev. B}\ }\textbf {\bibinfo {volume} {105}},\ \bibinfo {pages} {195140} (\bibinfo {year} {2022})}\BibitemShut {NoStop}%
\bibitem [{\citenamefont {Corboz}\ \emph {et~al.}(2011)\citenamefont {Corboz}, \citenamefont {White}, \citenamefont {Vidal},\ and\ \citenamefont {Troyer}}]{corboz-stripes-2011}%
  \BibitemOpen
  \bibfield  {author} {\bibinfo {author} {\bibfnamefont {P.}~\bibnamefont {Corboz}}, \bibinfo {author} {\bibfnamefont {S.~R.}\ \bibnamefont {White}}, \bibinfo {author} {\bibfnamefont {G.}~\bibnamefont {Vidal}},\ and\ \bibinfo {author} {\bibfnamefont {M.}~\bibnamefont {Troyer}},\ }\bibfield  {title} {\bibinfo {title} {Stripes in the two-dimensional $t$-$j$ model with infinite projected entangled-pair states},\ }\href {https://doi.org/10.1103/PhysRevB.84.041108} {\bibfield  {journal} {\bibinfo  {journal} {Phys. Rev. B}\ }\textbf {\bibinfo {volume} {84}},\ \bibinfo {pages} {041108} (\bibinfo {year} {2011})}\BibitemShut {NoStop}%
\bibitem [{\citenamefont {Huang}(2011{\natexlab{a}})}]{huang-biorthonormal-2011-b}%
  \BibitemOpen
  \bibfield  {author} {\bibinfo {author} {\bibfnamefont {Y.-K.}\ \bibnamefont {Huang}},\ }\bibfield  {title} {\bibinfo {title} {Biorthonormal transfer-matrix renormalization-group method for non-hermitian matrices},\ }\href {https://doi.org/10.1103/PhysRevE.83.036702} {\bibfield  {journal} {\bibinfo  {journal} {Phys. Rev. E}\ }\textbf {\bibinfo {volume} {83}},\ \bibinfo {pages} {036702} (\bibinfo {year} {2011}{\natexlab{a}})}\BibitemShut {NoStop}%
\bibitem [{\citenamefont {Huang}(2011{\natexlab{b}})}]{huang-biorthonormal-2011}%
  \BibitemOpen
  \bibfield  {author} {\bibinfo {author} {\bibfnamefont {Y.-K.}\ \bibnamefont {Huang}},\ }\bibfield  {title} {\bibinfo {title} {Biorthonormal matrix-product-state analysis for the non-hermitian transfer-matrix renormalization group in the thermodynamic limit},\ }\href {https://iopscience.iop.org/article/10.1088/1742-5468/2011/07/P07003} {\bibfield  {journal} {\bibinfo  {journal} {J. Stat. Mech. Theory Exp.}\ }\textbf {\bibinfo {volume} {2011}},\ \bibinfo {pages} {P07003} (\bibinfo {year} {2011}{\natexlab{b}})}\BibitemShut {NoStop}%
\bibitem [{\citenamefont {Huang}\ \emph {et~al.}(2012)\citenamefont {Huang}, \citenamefont {Chen},\ and\ \citenamefont {Kao}}]{huang-accurate-2012}%
  \BibitemOpen
  \bibfield  {author} {\bibinfo {author} {\bibfnamefont {Y.-K.}\ \bibnamefont {Huang}}, \bibinfo {author} {\bibfnamefont {P.}~\bibnamefont {Chen}},\ and\ \bibinfo {author} {\bibfnamefont {Y.-J.}\ \bibnamefont {Kao}},\ }\bibfield  {title} {\bibinfo {title} {Accurate computation of low-temperature thermodynamics for quantum spin chains},\ }\href {https://doi.org/10.1103/PhysRevB.86.235102} {\bibfield  {journal} {\bibinfo  {journal} {Phys. Rev. B}\ }\textbf {\bibinfo {volume} {86}},\ \bibinfo {pages} {235102} (\bibinfo {year} {2012})}\BibitemShut {NoStop}%
\bibitem [{\citenamefont {Fishman}\ \emph {et~al.}(2018)\citenamefont {Fishman}, \citenamefont {Vanderstraeten}, \citenamefont {Zauner-Stauber}, \citenamefont {Haegeman},\ and\ \citenamefont {Verstraete}}]{fishman-faster-2018}%
  \BibitemOpen
  \bibfield  {author} {\bibinfo {author} {\bibfnamefont {M.~T.}\ \bibnamefont {Fishman}}, \bibinfo {author} {\bibfnamefont {L.}~\bibnamefont {Vanderstraeten}}, \bibinfo {author} {\bibfnamefont {V.}~\bibnamefont {Zauner-Stauber}}, \bibinfo {author} {\bibfnamefont {J.}~\bibnamefont {Haegeman}},\ and\ \bibinfo {author} {\bibfnamefont {F.}~\bibnamefont {Verstraete}},\ }\bibfield  {title} {\bibinfo {title} {Faster methods for contracting infinite two-dimensional tensor networks},\ }\href {https://doi.org/10.1103/PhysRevB.98.235148} {\bibfield  {journal} {\bibinfo  {journal} {Phys. Rev. B}\ }\textbf {\bibinfo {volume} {98}},\ \bibinfo {pages} {235148} (\bibinfo {year} {2018})}\BibitemShut {NoStop}%
\bibitem [{\citenamefont {Acuaviva}\ \emph {et~al.}(2022)\citenamefont {Acuaviva}, \citenamefont {Makam}, \citenamefont {Nieuwboer}, \citenamefont {P{\'e}rez-Garc{\'\i}a}, \citenamefont {Sittner}, \citenamefont {Walter},\ and\ \citenamefont {Witteveen}}]{acuaviva-minimal-2022}%
  \BibitemOpen
  \bibfield  {author} {\bibinfo {author} {\bibfnamefont {A.}~\bibnamefont {Acuaviva}}, \bibinfo {author} {\bibfnamefont {V.}~\bibnamefont {Makam}}, \bibinfo {author} {\bibfnamefont {H.}~\bibnamefont {Nieuwboer}}, \bibinfo {author} {\bibfnamefont {D.}~\bibnamefont {P{\'e}rez-Garc{\'\i}a}}, \bibinfo {author} {\bibfnamefont {F.}~\bibnamefont {Sittner}}, \bibinfo {author} {\bibfnamefont {M.}~\bibnamefont {Walter}},\ and\ \bibinfo {author} {\bibfnamefont {F.}~\bibnamefont {Witteveen}},\ }\bibfield  {title} {\bibinfo {title} {The minimal canonical form of a tensor network},\ }\href {https://arxiv.org/abs/2209.14358} {\bibfield  {journal} {\bibinfo  {journal} {arXiv:2209.14358}\ } (\bibinfo {year} {2022})}\BibitemShut {NoStop}%
\bibitem [{\citenamefont {Vanhecke}\ \emph {et~al.}(2021{\natexlab{a}})\citenamefont {Vanhecke}, \citenamefont {Van~Damme}, \citenamefont {Haegeman}, \citenamefont {Vanderstraeten},\ and\ \citenamefont {Verstraete}}]{vanhecke-tangent-2021}%
  \BibitemOpen
  \bibfield  {author} {\bibinfo {author} {\bibfnamefont {B.}~\bibnamefont {Vanhecke}}, \bibinfo {author} {\bibfnamefont {M.}~\bibnamefont {Van~Damme}}, \bibinfo {author} {\bibfnamefont {J.}~\bibnamefont {Haegeman}}, \bibinfo {author} {\bibfnamefont {L.}~\bibnamefont {Vanderstraeten}},\ and\ \bibinfo {author} {\bibfnamefont {F.}~\bibnamefont {Verstraete}},\ }\bibfield  {title} {\bibinfo {title} {Tangent-space methods for truncating uniform mps},\ }\href {https://doi.org/10.21468/SciPostPhysCore.4.1.004} {\bibfield  {journal} {\bibinfo  {journal} {SciPost Phys. Core}\ }\textbf {\bibinfo {volume} {4}},\ \bibinfo {pages} {004} (\bibinfo {year} {2021}{\natexlab{a}})}\BibitemShut {NoStop}%
\bibitem [{\citenamefont {Brody}(2013)}]{brody-biorthogonal-2013}%
  \BibitemOpen
  \bibfield  {author} {\bibinfo {author} {\bibfnamefont {D.~C.}\ \bibnamefont {Brody}},\ }\bibfield  {title} {\bibinfo {title} {Biorthogonal quantum mechanics},\ }\href {https://iopscience.iop.org/article/10.1088/1751-8113/47/3/035305} {\bibfield  {journal} {\bibinfo  {journal} {J. Phys. A: Math.}\ }\textbf {\bibinfo {volume} {47}},\ \bibinfo {pages} {035305} (\bibinfo {year} {2013})}\BibitemShut {NoStop}%
\bibitem [{\citenamefont {Cardy}(1985)}]{cardy-conformal-1985}%
  \BibitemOpen
  \bibfield  {author} {\bibinfo {author} {\bibfnamefont {J.~L.}\ \bibnamefont {Cardy}},\ }\bibfield  {title} {\bibinfo {title} {Conformal invariance and the yang-lee edge singularity in two dimensions},\ }\href {https://doi.org/10.1103/PhysRevLett.54.1354} {\bibfield  {journal} {\bibinfo  {journal} {Phys. Rev. Lett.}\ }\textbf {\bibinfo {volume} {54}},\ \bibinfo {pages} {1354} (\bibinfo {year} {1985})}\BibitemShut {NoStop}%
\bibitem [{\citenamefont {Bianchini}\ \emph {et~al.}(2014)\citenamefont {Bianchini}, \citenamefont {Castro-Alvaredo}, \citenamefont {Doyon}, \citenamefont {Levi},\ and\ \citenamefont {Ravanini}}]{bianchini-entanglement-2014}%
  \BibitemOpen
  \bibfield  {author} {\bibinfo {author} {\bibfnamefont {D.}~\bibnamefont {Bianchini}}, \bibinfo {author} {\bibfnamefont {O.}~\bibnamefont {Castro-Alvaredo}}, \bibinfo {author} {\bibfnamefont {B.}~\bibnamefont {Doyon}}, \bibinfo {author} {\bibfnamefont {E.}~\bibnamefont {Levi}},\ and\ \bibinfo {author} {\bibfnamefont {F.}~\bibnamefont {Ravanini}},\ }\bibfield  {title} {\bibinfo {title} {Entanglement entropy of non-unitary conformal field theory},\ }\href {https://iopscience.iop.org/article/10.1088/1751-8113/48/4/04FT01} {\bibfield  {journal} {\bibinfo  {journal} {J. Phys. A: Math. Theor.}\ }\textbf {\bibinfo {volume} {48}},\ \bibinfo {pages} {04FT01} (\bibinfo {year} {2014})}\BibitemShut {NoStop}%
\bibitem [{\citenamefont {Bianchini}\ \emph {et~al.}(2015)\citenamefont {Bianchini}, \citenamefont {Castro-Alvaredo},\ and\ \citenamefont {Doyon}}]{bianchini-entanglement-2015}%
  \BibitemOpen
  \bibfield  {author} {\bibinfo {author} {\bibfnamefont {D.}~\bibnamefont {Bianchini}}, \bibinfo {author} {\bibfnamefont {O.~A.}\ \bibnamefont {Castro-Alvaredo}},\ and\ \bibinfo {author} {\bibfnamefont {B.}~\bibnamefont {Doyon}},\ }\bibfield  {title} {\bibinfo {title} {Entanglement entropy of non-unitary integrable quantum field theory},\ }\href {https://www.sciencedirect.com/science/article/pii/S055032131500173X?via%3Dihub} {\bibfield  {journal} {\bibinfo  {journal} {Nucl. Phys. B}\ }\textbf {\bibinfo {volume} {896}},\ \bibinfo {pages} {835} (\bibinfo {year} {2015})}\BibitemShut {NoStop}%
\bibitem [{\citenamefont {Bianchini}\ and\ \citenamefont {Ravanini}(2016)}]{bianchini-entanglement-2016}%
  \BibitemOpen
  \bibfield  {author} {\bibinfo {author} {\bibfnamefont {D.}~\bibnamefont {Bianchini}}\ and\ \bibinfo {author} {\bibfnamefont {F.}~\bibnamefont {Ravanini}},\ }\bibfield  {title} {\bibinfo {title} {Entanglement entropy from corner transfer matrix in forrester--baxter non-unitary rsos models},\ }\href {https://iopscience.iop.org/article/10.1088/1751-8113/49/15/154005} {\bibfield  {journal} {\bibinfo  {journal} {J. Phys. A: Math. Theor.}\ }\textbf {\bibinfo {volume} {49}},\ \bibinfo {pages} {154005} (\bibinfo {year} {2016})}\BibitemShut {NoStop}%
\bibitem [{\citenamefont {Couvreur}\ \emph {et~al.}(2017)\citenamefont {Couvreur}, \citenamefont {Jacobsen},\ and\ \citenamefont {Saleur}}]{couvreur-entanglement-2017}%
  \BibitemOpen
  \bibfield  {author} {\bibinfo {author} {\bibfnamefont {R.}~\bibnamefont {Couvreur}}, \bibinfo {author} {\bibfnamefont {J.~L.}\ \bibnamefont {Jacobsen}},\ and\ \bibinfo {author} {\bibfnamefont {H.}~\bibnamefont {Saleur}},\ }\bibfield  {title} {\bibinfo {title} {Entanglement in nonunitary quantum critical spin chains},\ }\href {https://doi.org/10.1103/PhysRevLett.119.040601} {\bibfield  {journal} {\bibinfo  {journal} {Phys. Rev. Lett.}\ }\textbf {\bibinfo {volume} {119}},\ \bibinfo {pages} {040601} (\bibinfo {year} {2017})}\BibitemShut {NoStop}%
\bibitem [{\citenamefont {Dupic}\ \emph {et~al.}(2018)\citenamefont {Dupic}, \citenamefont {Estienne},\ and\ \citenamefont {Ikhlef}}]{dupic-entanglement-2018}%
  \BibitemOpen
  \bibfield  {author} {\bibinfo {author} {\bibfnamefont {T.}~\bibnamefont {Dupic}}, \bibinfo {author} {\bibfnamefont {B.}~\bibnamefont {Estienne}},\ and\ \bibinfo {author} {\bibfnamefont {Y.}~\bibnamefont {Ikhlef}},\ }\bibfield  {title} {\bibinfo {title} {Entanglement entropies of minimal models from null-vectors},\ }\href {https://doi.org/10.21468/SciPostPhys.4.6.031} {\bibfield  {journal} {\bibinfo  {journal} {SciPost Phys.}\ }\textbf {\bibinfo {volume} {4}},\ \bibinfo {pages} {031} (\bibinfo {year} {2018})}\BibitemShut {NoStop}%
\bibitem [{\citenamefont {Tu}\ \emph {et~al.}(2022)\citenamefont {Tu}, \citenamefont {Tzeng},\ and\ \citenamefont {Chang}}]{tu-renyi-2022}%
  \BibitemOpen
  \bibfield  {author} {\bibinfo {author} {\bibfnamefont {Y.-T.}\ \bibnamefont {Tu}}, \bibinfo {author} {\bibfnamefont {Y.-C.}\ \bibnamefont {Tzeng}},\ and\ \bibinfo {author} {\bibfnamefont {P.-Y.}\ \bibnamefont {Chang}},\ }\bibfield  {title} {\bibinfo {title} {R{\'e}nyi entropies and negative central charges in non-hermitian quantum systems},\ }\href {https://doi.org/10.21468/SciPostPhys.12.6.194} {\bibfield  {journal} {\bibinfo  {journal} {SciPost Phys.}\ }\textbf {\bibinfo {volume} {12}},\ \bibinfo {pages} {194} (\bibinfo {year} {2022})}\BibitemShut {NoStop}%
\bibitem [{\citenamefont {Pollmann}\ \emph {et~al.}(2009)\citenamefont {Pollmann}, \citenamefont {Mukerjee}, \citenamefont {Turner},\ and\ \citenamefont {Moore}}]{pollmann-theory-2009}%
  \BibitemOpen
  \bibfield  {author} {\bibinfo {author} {\bibfnamefont {F.}~\bibnamefont {Pollmann}}, \bibinfo {author} {\bibfnamefont {S.}~\bibnamefont {Mukerjee}}, \bibinfo {author} {\bibfnamefont {A.~M.}\ \bibnamefont {Turner}},\ and\ \bibinfo {author} {\bibfnamefont {J.~E.}\ \bibnamefont {Moore}},\ }\bibfield  {title} {\bibinfo {title} {Theory of finite-entanglement scaling at one-dimensional quantum critical points},\ }\href {https://doi.org/10.1103/PhysRevLett.102.255701} {\bibfield  {journal} {\bibinfo  {journal} {Phys. Rev. Lett.}\ }\textbf {\bibinfo {volume} {102}},\ \bibinfo {pages} {255701} (\bibinfo {year} {2009})}\BibitemShut {NoStop}%
\bibitem [{\citenamefont {Nuomin}\ \emph {et~al.}(2022)\citenamefont {Nuomin}, \citenamefont {Song}, \citenamefont {Beratan},\ and\ \citenamefont {Zhang}}]{PhysRevB.106.104306}%
  \BibitemOpen
  \bibfield  {author} {\bibinfo {author} {\bibfnamefont {H.}~\bibnamefont {Nuomin}}, \bibinfo {author} {\bibfnamefont {F.-F.}\ \bibnamefont {Song}}, \bibinfo {author} {\bibfnamefont {D.~N.}\ \bibnamefont {Beratan}},\ and\ \bibinfo {author} {\bibfnamefont {P.}~\bibnamefont {Zhang}},\ }\bibfield  {title} {\bibinfo {title} {Suppressing the entanglement growth in matrix product state evolution of quantum systems through nonunitary similarity transformations},\ }\href {https://doi.org/10.1103/PhysRevB.106.104306} {\bibfield  {journal} {\bibinfo  {journal} {Phys. Rev. B}\ }\textbf {\bibinfo {volume} {106}},\ \bibinfo {pages} {104306} (\bibinfo {year} {2022})}\BibitemShut {NoStop}%
\bibitem [{\citenamefont {McCulloch}(2008)}]{mcculloch-infinite-2008}%
  \BibitemOpen
  \bibfield  {author} {\bibinfo {author} {\bibfnamefont {I.~P.}\ \bibnamefont {McCulloch}},\ }\bibfield  {title} {\bibinfo {title} {Infinite size density matrix renormalization group, revisited},\ }\href {https://arxiv.org/abs/0804.2509v1} {\bibfield  {journal} {\bibinfo  {journal} {arXiv:0804.2509}\ } (\bibinfo {year} {2008})}\BibitemShut {NoStop}%
\bibitem [{Note1()}]{Note1}%
  \BibitemOpen
  \bibinfo {note} {For a matrix $M$, consider $M_\epsilon $ = $\protect \mathrm {e}^{-\epsilon Q} M \protect \mathrm {e}^{\epsilon Q}$, where $\epsilon \ll 1$ and $Q$ is a Hermitian matrix. Then $\protect \mathrm {Tr}(M_\epsilon ^\dagger M_\epsilon ) = \protect \mathrm {Tr}(M^\dagger M) + 2\epsilon \protect \mathrm {Tr}(Q [M^\dagger , M]) + 2 \epsilon ^2 \protect \mathrm {Tr}([M, Q]^\dagger [M, Q])$. If the matrix $M$ is normal, then $[M^\dagger , M] = 0$ and the first order term vanishes. As the second order term is positive, this indicates that the point where $M$ is normal is a minimum.}\BibitemShut {Stop}%
\bibitem [{\citenamefont {Wannier}(1950)}]{wannier-antiferromagnetism-1950}%
  \BibitemOpen
  \bibfield  {author} {\bibinfo {author} {\bibfnamefont {G.~H.}\ \bibnamefont {Wannier}},\ }\bibfield  {title} {\bibinfo {title} {Antiferromagnetism. the triangular ising net},\ }\href {https://doi.org/10.1103/PhysRev.79.357} {\bibfield  {journal} {\bibinfo  {journal} {Phys. Rev.}\ }\textbf {\bibinfo {volume} {79}},\ \bibinfo {pages} {357} (\bibinfo {year} {1950})}\BibitemShut {NoStop}%
\bibitem [{\citenamefont {Vanhecke}\ \emph {et~al.}(2021{\natexlab{b}})\citenamefont {Vanhecke}, \citenamefont {Colbois}, \citenamefont {Vanderstraeten}, \citenamefont {Verstraete},\ and\ \citenamefont {Mila}}]{vanhecke-solving-2021}%
  \BibitemOpen
  \bibfield  {author} {\bibinfo {author} {\bibfnamefont {B.}~\bibnamefont {Vanhecke}}, \bibinfo {author} {\bibfnamefont {J.}~\bibnamefont {Colbois}}, \bibinfo {author} {\bibfnamefont {L.}~\bibnamefont {Vanderstraeten}}, \bibinfo {author} {\bibfnamefont {F.}~\bibnamefont {Verstraete}},\ and\ \bibinfo {author} {\bibfnamefont {F.}~\bibnamefont {Mila}},\ }\bibfield  {title} {\bibinfo {title} {Solving frustrated ising models using tensor networks},\ }\href {https://doi.org/10.1103/PhysRevResearch.3.013041} {\bibfield  {journal} {\bibinfo  {journal} {Phys. Rev. Research}\ }\textbf {\bibinfo {volume} {3}},\ \bibinfo {pages} {013041} (\bibinfo {year} {2021}{\natexlab{b}})}\BibitemShut {NoStop}%
\bibitem [{\citenamefont {Nourhani}\ \emph {et~al.}(2018)\citenamefont {Nourhani}, \citenamefont {Crespi},\ and\ \citenamefont {Lammert}}]{nourhani-communicating-2018}%
  \BibitemOpen
  \bibfield  {author} {\bibinfo {author} {\bibfnamefont {A.}~\bibnamefont {Nourhani}}, \bibinfo {author} {\bibfnamefont {V.~H.}\ \bibnamefont {Crespi}},\ and\ \bibinfo {author} {\bibfnamefont {P.~E.}\ \bibnamefont {Lammert}},\ }\bibfield  {title} {\bibinfo {title} {Communicating through a sea of frustration: Zero-temperature triangular ising antiferromagnet on a cylinder},\ }\href {https://doi.org/10.1103/PhysRevE.98.032107} {\bibfield  {journal} {\bibinfo  {journal} {Phys. Rev. E}\ }\textbf {\bibinfo {volume} {98}},\ \bibinfo {pages} {032107} (\bibinfo {year} {2018})}\BibitemShut {NoStop}%
\bibitem [{\citenamefont {Bl\"ote}\ and\ \citenamefont {Nightingale}(1993)}]{blote-antiferromagnetic-1993}%
  \BibitemOpen
  \bibfield  {author} {\bibinfo {author} {\bibfnamefont {H.~W.~J.}\ \bibnamefont {Bl\"ote}}\ and\ \bibinfo {author} {\bibfnamefont {M.~P.}\ \bibnamefont {Nightingale}},\ }\bibfield  {title} {\bibinfo {title} {Antiferromagnetic triangular ising model: Critical behavior of the ground state},\ }\href {https://doi.org/10.1103/PhysRevB.47.15046} {\bibfield  {journal} {\bibinfo  {journal} {Phys. Rev. B}\ }\textbf {\bibinfo {volume} {47}},\ \bibinfo {pages} {15046} (\bibinfo {year} {1993})}\BibitemShut {NoStop}%
\bibitem [{\citenamefont {Jiang}\ and\ \citenamefont {Emig}(2006)}]{jiang-ordering-2006}%
  \BibitemOpen
  \bibfield  {author} {\bibinfo {author} {\bibfnamefont {Y.}~\bibnamefont {Jiang}}\ and\ \bibinfo {author} {\bibfnamefont {T.}~\bibnamefont {Emig}},\ }\bibfield  {title} {\bibinfo {title} {Ordering of geometrically frustrated classical and quantum triangular ising magnets},\ }\href {https://doi.org/10.1103/PhysRevB.73.104452} {\bibfield  {journal} {\bibinfo  {journal} {Phys. Rev. B}\ }\textbf {\bibinfo {volume} {73}},\ \bibinfo {pages} {104452} (\bibinfo {year} {2006})}\BibitemShut {NoStop}%
\bibitem [{\citenamefont {Smerald}\ and\ \citenamefont {Mila}(2018)}]{smerald-spinliquid-2018}%
  \BibitemOpen
  \bibfield  {author} {\bibinfo {author} {\bibfnamefont {A.}~\bibnamefont {Smerald}}\ and\ \bibinfo {author} {\bibfnamefont {F.}~\bibnamefont {Mila}},\ }\bibfield  {title} {\bibinfo {title} {Spin-liquid behaviour and the interplay between pokrovsky-talapov and ising criticality in the distorted, triangular-lattice, dipolar ising antiferromagnet},\ }\href {https://scipost.org/10.21468/SciPostPhys.5.3.030} {\bibfield  {journal} {\bibinfo  {journal} {SciPost Phys.}\ }\textbf {\bibinfo {volume} {5}},\ \bibinfo {pages} {030} (\bibinfo {year} {2018})}\BibitemShut {NoStop}%
\bibitem [{\citenamefont {Jiang}\ and\ \citenamefont {Ran}(2015)}]{jiang-symmetric-2015}%
  \BibitemOpen
  \bibfield  {author} {\bibinfo {author} {\bibfnamefont {S.}~\bibnamefont {Jiang}}\ and\ \bibinfo {author} {\bibfnamefont {Y.}~\bibnamefont {Ran}},\ }\bibfield  {title} {\bibinfo {title} {Symmetric tensor networks and practical simulation algorithms to sharply identify classes of quantum phases distinguishable by short-range physics},\ }\href {https://doi.org/10.1103/PhysRevB.92.104414} {\bibfield  {journal} {\bibinfo  {journal} {Phys. Rev. B}\ }\textbf {\bibinfo {volume} {92}},\ \bibinfo {pages} {104414} (\bibinfo {year} {2015})}\BibitemShut {NoStop}%
\bibitem [{\citenamefont {Mambrini}\ \emph {et~al.}(2016)\citenamefont {Mambrini}, \citenamefont {Or\'us},\ and\ \citenamefont {Poilblanc}}]{mambrini-systematic-2016}%
  \BibitemOpen
  \bibfield  {author} {\bibinfo {author} {\bibfnamefont {M.}~\bibnamefont {Mambrini}}, \bibinfo {author} {\bibfnamefont {R.}~\bibnamefont {Or\'us}},\ and\ \bibinfo {author} {\bibfnamefont {D.}~\bibnamefont {Poilblanc}},\ }\bibfield  {title} {\bibinfo {title} {Systematic construction of spin liquids on the square lattice from tensor networks with su(2) symmetry},\ }\href {https://doi.org/10.1103/PhysRevB.94.205124} {\bibfield  {journal} {\bibinfo  {journal} {Phys. Rev. B}\ }\textbf {\bibinfo {volume} {94}},\ \bibinfo {pages} {205124} (\bibinfo {year} {2016})}\BibitemShut {NoStop}%
\bibitem [{\citenamefont {Hackenbroich}\ \emph {et~al.}(2018)\citenamefont {Hackenbroich}, \citenamefont {Sterdyniak},\ and\ \citenamefont {Schuch}}]{hackenbroich-interplay-2018}%
  \BibitemOpen
  \bibfield  {author} {\bibinfo {author} {\bibfnamefont {A.}~\bibnamefont {Hackenbroich}}, \bibinfo {author} {\bibfnamefont {A.}~\bibnamefont {Sterdyniak}},\ and\ \bibinfo {author} {\bibfnamefont {N.}~\bibnamefont {Schuch}},\ }\bibfield  {title} {\bibinfo {title} {Interplay of su(2), point group, and translational symmetry for projected entangled pair states: Application to a chiral spin liquid},\ }\href {https://doi.org/10.1103/PhysRevB.98.085151} {\bibfield  {journal} {\bibinfo  {journal} {Phys. Rev. B}\ }\textbf {\bibinfo {volume} {98}},\ \bibinfo {pages} {085151} (\bibinfo {year} {2018})}\BibitemShut {NoStop}%
\bibitem [{\citenamefont {Van~Damme}\ \emph {et~al.}(2024)\citenamefont {Van~Damme}, \citenamefont {Devos},\ and\ \citenamefont {Haegeman}}]{Van_Damme_MPSKit_2024}%
  \BibitemOpen
  \bibfield  {author} {\bibinfo {author} {\bibfnamefont {M.}~\bibnamefont {Van~Damme}}, \bibinfo {author} {\bibfnamefont {L.}~\bibnamefont {Devos}},\ and\ \bibinfo {author} {\bibfnamefont {J.}~\bibnamefont {Haegeman}},\ }\href {https://doi.org/10.5281/zenodo.10654901} {\bibinfo {title} {{MPSKit}}} (\bibinfo {year} {2024})\BibitemShut {NoStop}%
\end{thebibliography}%

\end{document}